\pdfoutput=1
\RequirePackage{ifpdf}
\ifpdf
\documentclass[pdftex]{sigma}
\else
\documentclass{sigma}
\fi

\numberwithin{equation}{section}

\usepackage[ps,arrow,matrix]{xy} 

\newcommand{\im}{\mathrm{i}}
\newcommand{\N}{\mathbb{N}}
\newcommand{\Q}{\mathbb{Q}}
\newcommand{\R}{\mathbb{R}}
\newcommand{\C}{\mathbb{C}}
\newcommand{\K}{\mathbb{K}}
\newcommand{\defeq}{:=}
\newcommand{\tens}{\otimes}
\newcommand{\ctens}{\hat{\otimes}}
\newcommand{\cH}{\mathcal{H}}
\newcommand{\rH}{\mathrm{H}}
\newcommand{\cL}{\mathcal{L}}
\newcommand{\xd}{\mathrm{d}}

\newcommand{\toi}{\hookrightarrow}
\newcommand{\tos}{\twoheadrightarrow}
\newcommand{\id}{\mathrm{id}}
\newcommand{\cM}{\mathcal{M}}
\newcommand{\cN}{\mathcal{N}}
\newcommand{\one}{\mathbf{1}}
\newcommand{\ds}{\circ}

\theoremstyle{definition}
\newtheorem{dfn}{Definition}[section]

\theoremstyle{plain}
\newtheorem{lem}[dfn]{Lemma}
\newtheorem{prop}[dfn]{Proposition}
\newtheorem{thm}[dfn]{Theorem}

\allowdisplaybreaks

\begin{document}

\renewcommand{\PaperNumber}{050}

\FirstPageHeading

\ShortArticleName{Holomorphic Quantization of Linear Field Theory in the General Boundary Formulation}

\ArticleName{Holomorphic Quantization of Linear Field Theory\\ in the General Boundary Formulation}

\Author{Robert OECKL}

\AuthorNameForHeading{R.~Oeckl}

\Address{Instituto de Matem\'aticas, Universidad Nacional Aut\'onoma de M\'exico,\\
Campus Morelia, C.P.~58190, Morelia, Michoac\'an, Mexico}

\Email{\href{mailto:robert@matmor.unam.mx}{robert@matmor.unam.mx}}
\URLaddress{\url{http://www.matmor.unam.mx/~robert/}}

\ArticleDates{Received April 27, 2012, in f\/inal form August 03, 2012; Published online August 09, 2012}

\Abstract{We present a rigorous quantization scheme that yields a quantum f\/ield theory in general boundary form starting from a linear f\/ield theory. Following a geometric quantization approach in the K\"ahler case, state spaces arise as spaces of holomorphic functions on linear spaces of classical solutions in neighborhoods of hypersurfaces. Amplitudes arise as integrals of such functions over spaces of classical solutions in regions of spacetime. We prove the validity of the TQFT-type axioms of the general boundary formulation under reasonable assumptions. We also develop the notions of vacuum and coherent states in this framework. As a f\/irst application we quantize evanescent waves in Klein--Gordon theory.}


\Classification{57R56; 81S10; 81T05; 81T20}

\section{Introduction}

The general boundary formulation (GBF) is an axiomatic approach to quantum theory and to quantum f\/ield theory in particular \cite{Oe:boundary,Oe:GBQFT}. Its mathematical structure consists of an axiomatic system in the spirit of topological quantum f\/ield theory (TQFT) \cite{Ati:tqft} and shows similarities with Segal's approach to quantum f\/ield theory \cite{Seg:cftdef}. Its physical interpretation is based on a probability rule that generalizes the Born rule \cite{Oe:GBQFT,Oe:probgbf}. Key features of the GBF are that it provides a strictly local description of physics and accommodates theories without metric background naturally. The latter feature is in marked contrast to older and more established axiomatic approaches to quantum f\/ield theory such as those based on the Wightman axioms~\cite{StWi:pct}, the Osterwalder--Schrader axioms \cite{OsSc:axeucl,OsSc:axeucl2} or the Haag--Kastler axioms~\cite{HaKa:aqft}. The GBF was conceived particularly as a possible framework to accommodate a quantum theory of gravity \cite{CDORT:vacuum,Oe:boundary,Oe:GBQFT,Oe:catandclock,Oe:bqgrav}.

Quantum f\/ield theories in a GBF form have been described so far principally by using Feynman path integral quantization combined with the Schr\"odinger representation \cite{Col:vacuum,Col:desitterletter,CoOe:unitary,CoOe:spsmatrix,CoOe:smatrixgbf,CoOe:2deucl,CDORT:vacuum,CoRo:genschroed,Dop:tomschwing,Oe:boundary,Oe:KGtl,Oe:timelike,Oe:2dqym}. Compared to ``inf\/initesimal'' approaches that are based on exponentiating a generator of time evolution, such as canonical quantization, the path integral has the advantage of naturally generalizing at least formally to general spacetime regions which are neither naturally nor uniquely described by some kind of ``evolution''. But the Schr\"odinger--Feynman approach has also disadvantages. In particular, it is not in general clear how the Schr\"odinger representation should be def\/ined, the boundary value problems that occur are in general not well-posed and the Feynman path integral is frequently ill def\/ined. These problems can be solved or mitigated for particular theories and contexts (as shown for example by the mentioned works), but a quantization scheme that avoids these problems is clearly desirable.

The present article is concerned with a quantization scheme that starts with a linear f\/ield theory and yields a quantum f\/ield theory satisfying the axioms of the GBF. The classical theo\-ry is provided in the form of spaces of solutions of the f\/ield equations near hypersurfaces and in regions of spacetime. These spaces of solutions are real vector spaces, but carry additional structure in the case of hypersurfaces, making them into complex Hilbert spaces. The additional structure is a symplectic structure that can naturally be obtained from a Lagrangian formulation of the classical theory as well as a compatible complex structure. The quantization for each hypersurface proceeds then as a geometric quantization in the K\"ahler case, that is, as a~holomorphic quantization. The Hilbert spaces so obtained are Fock spaces, realized as spaces of holomorphic functions. The quantization for each spacetime region consists in the assignment to each holomorphic function on the boundary, of its integral over the space of classical solutions in the region's interior. This yields the amplitude maps.

In order to be able to properly formulate the quantization scheme it is necessary to use the theory of Gaussian integration in inf\/inite-dimensional vector spaces. While this is not in itself a new subject, we have not found in the literature a treatment suitable to our needs. We thus develop the relevant theory here ourselves. In particular, it is important that the we deal with a proper positive measure and that the arising Fock spaces are concretely realized as spaces of holomorphic functions in the spirit of Bargmann's approach~\cite{Bar:hilbanalytic,Bar:remanalytic}. This also facilitates a proper treatment of coherent states.

As a f\/irst application we quantize Klein--Gordon theory for three dif\/ferent choices of admissible hypersurfaces in Minkowski space. The f\/irst choice is the ``standard choice'' of equal-time hypersurfaces, provided for comparison. The other two choices involve families of timelike hypersurfaces that previously were treated in the GBF via Schr\"odinger--Feynman quantization \cite{CoOe:spsmatrix,CoOe:smatrixgbf,Oe:KGtl, Oe:timelike}. In contrast to those treatments, however, we provide a complete quantization of the relevant spaces of classical solutions. That is, the quantization not only includes propagating waves, but also evanescent waves, i.e., waves that behave exponentially in space.

We start in Section~\ref{sec:axioms} by providing a suitable version of the axiomatic framework of the GBF, with a description of the geometric data, the core axioms and the vacuum axioms. In Section~\ref{sec:ingredients} we review ingredients for the quantization, including aspects of Lagrangian f\/ield theory, geometric quantization as well as mathematical foundations of Gaussian integration and holomorphic functions in inf\/inite-dimensional vector spaces, as used subsequently. Section~\ref{sec:quantization} provides the core of the paper, laying out in detail the quantization scheme, the role of coherent states and some aspects of the physical interpretation. Section~\ref{sec:kg} presents the application to Klein--Gordon theory for three dif\/ferent types of geometric setting, incorporating a quantization of evanescent waves in the case of timelike hypersurfaces. Finally, we provide some conclusions and an outlook in Section~\ref{sec:outlook}.

{\bf Note added.}
Most of Section~\ref{sec:geomax} as well as part of Section~\ref{sec:coreaxioms} were published in essentially identical form in~\cite{Oeckl2012}
 as Section~3.1 and~3.3 respectively. However, before this publication by a~journal, this content was already publicly available in the preprint version of the present paper. 
 This is clearly referenced in the abovementioned paper. Moreover, this content is nothing but a slightly adapted version of material published earlier in the paper~\cite{Oe:GBQFT} by the author as the initial part of Section~2.

\section{Axioms}
\label{sec:axioms}

\subsection{Geometric data}
\label{sec:geomax}

\looseness=1
In the tradition of TQFT we may think of the core axioms of the GBF as describing an association of ``algebraic'' data to ``geometric'' data as well as properties that this association satisf\/ies. In the present case on the ``algebraic'' side we are mainly dealing with Hilbert spaces and maps between Hilbert spaces. (The description ``functional analytic'' data instead of ``algebraic'' data might be more appropriate.) On the ``geometric'' side we are dealing with certain topological manifolds, possibly with additional structure. In the following we will make use only of purely topological structure (and could even further abstract from that if desired). Nevertheless, we will continue to use the word ``geometric'' with the understanding that one should in general expect the topological manifolds to come equipped with additional structure.

Concretely, our geometric setting is the following: There is a f\/ixed positive integer $d\in\N$. We are given a collection of oriented topological manifolds of dimension $d$, possibly with boundary, that we call \emph{regions}. Furthermore, there is a collection of oriented topological manifolds without boundary of dimension $d-1$ that we call \emph{hypersurfaces}. All manifolds may only have f\/initely many connected components. When we want to emphasize explicitly that a given manifold is in one of those collections we also use the attribute \emph{admissible}. These collections satisfy the following requirements:
\begin{itemize}\itemsep=0pt
\item Any connected component of a region or hypersurface is admissible.
\item Any f\/inite disjoint union of regions or of hypersurfaces is admissible.
\item Any boundary of a region is an admissible hypersurface.
\item If $\Sigma$ is a hypersurface, then $\overline{\Sigma}$, denoting the same manifold with opposite orientation, is admissible.
\end{itemize}
It will turn out to be convenient to also introduce \emph{empty regions}. An empty region is topologically simply a hypersurface, but thought of as an inf\/initesimally thin region. Concretely, the empty region associated with a hypersurface $\Sigma$ will be denoted by $\hat{\Sigma}$ and its boundary is def\/ined to be the disjoint union $\partial \hat{\Sigma}=\Sigma\cup\overline{\Sigma}$. There is one empty region for each hypersurface (forgetting its orientation). When an explicit distinction is desirable we refer to the previously def\/ined regions as \emph{regular regions}.

There is also a notion of \emph{gluing} of regions. Suppose we are given a region $M$ with its boundary a disjoint union $\partial M=\Sigma_1\cup\Sigma\cup\overline{\Sigma'}$, where $\Sigma'$ is a copy of $\Sigma$. ($\Sigma_1$ may be empty.) Then, we may obtain a new manifold $M_1$ by gluing $M$ to itself along $\Sigma$, $\overline{\Sigma'}$. That is, we identify the points of $\Sigma$ with corresponding points of $\Sigma'$ to obtain $M_1$. The resulting manifold $M_1$ might be inadmissible, in which case the gluing is not allowed. The gluing is unique in the sense that if there are dif\/ferent ways to make this identif\/ication the resulting manifolds $M_1$ must be indistinguishable in our setting.

As already mentioned, the manifolds carry additional structure in general. This has to be taken into account in the gluing and will modify the procedure as well as its possibility in the f\/irst place. Our description above is merely meant as a minimal one. Moreover, there might be important information present in dif\/ferent ways of identifying the boundary hypersurfaces that are glued. Such a case can be incorporated into our present setting by encoding this information explicitly through suitable additional structure on the manifolds.

\subsection{Core axioms}
\label{sec:coreaxioms}

\looseness=1
The core axioms listed in the following are a ref\/inement of the axioms introduced in~\cite{Oe:GBQFT} and~\cite{Oe:2dqym}. In fact, the formulation here is closer to that in~\cite{Oe:2dqym}, but without corners. For easier compari\-son, we preserve to a large extend the naming conventions used in those works. Here, as in the following, $\tens$ denotes the tensor product of vector spaces, while $\ctens$ denotes the completed tensor product of Hilbert spaces. We add some more specif\/ic comments after the listing of the axioms.
\begin{itemize}\itemsep=0pt
\item[(T1)] Associated to each hypersurface $\Sigma$ is a complex
  separable Hilbert space $\cH_\Sigma$, called the \emph{state space} of
  $\Sigma$. We denote its inner product by
  $\langle\cdot,\cdot\rangle_\Sigma$.
\item[(T1b)] Associated to each hypersurface $\Sigma$ is a conjugate linear
  isometry $\iota_\Sigma:\cH_\Sigma\to\cH_{\overline{\Sigma}}$. This map
  is an involution in the sense that $\iota_{\overline{\Sigma}}\circ\iota_\Sigma$
  is the identity on  $\cH_\Sigma$.
\item[(T2)] Suppose the hypersurface $\Sigma$ decomposes into a disjoint
  union of hypersurfaces \mbox{$\Sigma=\Sigma_1\cup\cdots$} $\cdots\cup\Sigma_n$. Then,
  there is an isometric isomorphism of Hilbert spaces
  \mbox{$\tau_{\Sigma_1,\dots,\Sigma_n;\Sigma}:\cH_{\Sigma_1}\ctens\cdots$} $\cdots\ctens\cH_{\Sigma_n}\to\cH_\Sigma$.
  The composition of the maps $\tau$ associated with two consecutive
  decompositions is identical to the map $\tau$ associated to the
  resulting decomposition.
\item[(T2b)] The involution $\iota$ is compatible with the above
  decomposition. That is
  $\tau_{\overline{\Sigma}_1,\dots,\overline{\Sigma}_n;\overline{\Sigma}}
  \circ(\iota_{\Sigma_1}\ctens\cdots$ $\cdots\ctens\iota_{\Sigma_n})
  =\iota_\Sigma\circ\tau_{\Sigma_1,\dots,\Sigma_n;\Sigma}$.
\item[(T4)] Associated with each region $M$ is a linear map
  from a dense subspace $\cH_{\partial M}^\ds$ of the state space
  $\cH_{\partial M}$ of its boundary $\partial M$ (which carries the
  induced orientation) to the complex
  numbers, $\rho_M:\cH_{\partial M}^\ds\to\C$. This is called the
  \emph{amplitude} map.
\item[(T3x)] Let $\Sigma$ be a hypersurface. The boundary $\partial\hat{\Sigma}$ of the associated empty region $\hat{\Sigma}$ decomposes into the disjoint union $\partial\hat{\Sigma}=\overline{\Sigma}\cup\Sigma'$, where $\Sigma'$ denotes a second copy of $\Sigma$. Then, $\tau_{\overline{\Sigma},\Sigma';\partial\hat{\Sigma}}(\cH_{\overline{\Sigma}}\tens\cH_{\Sigma'})\subseteq\cH_{\partial\hat{\Sigma}}^\ds$. Moreover, $\rho_{\hat{\Sigma}}\circ\tau_{\overline{\Sigma},\Sigma';\partial\hat{\Sigma}}$ restricts to a bilinear pairing $(\cdot,\cdot)_\Sigma:\cH_{\overline{\Sigma}}\times\cH_{\Sigma'}\to\C$ such that $\langle\cdot,\cdot\rangle_\Sigma=(\iota_\Sigma(\cdot),\cdot)_\Sigma$.
\item[(T5a)] Let $M_1$ and $M_2$ be regions and $M\defeq M_1\cup M_2$ be their disjoint union\footnote{In the setting of a global background (compare terminology of \cite{Oe:GBQFT}), that is if all regions and hypersurfaces are submanifolds of a given manifold of dimension $d$, we may need to allow here unions that are disjoint only up to boundaries. However, in such a case we shall still think of the regions as ``disjoint''.}. Then $\partial M=\partial M_1\cup \partial M_2$ is also a disjoint union and $\tau_{\partial M_1,\partial M_2;\partial M}(\cH_{\partial M_1}^\ds\tens \cH_{\partial M_2}^\ds)\subseteq \cH_{\partial M}^\ds$. Moreover, for all $\psi_1\in\cH_{\partial M_1}^\ds$ and $\psi_2\in\cH_{\partial M_2}^\ds$
\begin{gather*}
 \rho_{M}\circ\tau_{\partial M_1,\partial M_2;\partial M}(\psi_1\tens\psi_2)= \rho_{M_1}(\psi_1)\rho_{M_2}(\psi_2) .
\end{gather*}
\item[(T5b)] Let $M$ be a region with its boundary decomposing as a disjoint union $\partial M=\Sigma_1\cup\Sigma\cup \overline{\Sigma'}$, where $\Sigma'$ is a copy of $\Sigma$. Let $M_1$ denote the gluing of $M$ with itself along $\Sigma,\overline{\Sigma'}$ and suppose that $M_1$ is a region. Note $\partial M_1=\Sigma_1$. Then, $\tau_{\Sigma_1,\Sigma,\overline{\Sigma'};\partial M}(\psi\tens\xi\tens\iota_\Sigma(\xi))\in\cH_{\partial M}^\ds$ for all $\psi\in\cH_{\partial M_1}^\ds$ and $\xi\in\cH_\Sigma$. Moreover, for any ON-basis $\{\xi_i\}_{i\in I}$ of $\cH_\Sigma$, we have for all $\psi\in\cH_{\partial M_1}^\ds$
\begin{gather}
 \rho_{M_1}(\psi)\cdot c(M;\Sigma,\overline{\Sigma'})
 =\sum_{i\in I}\rho_M\circ\tau_{\Sigma_1,\Sigma,\overline{\Sigma'};\partial M}(\psi\tens\xi_i\tens\iota_\Sigma(\xi_i)),
\label{eq:glueax1}
\end{gather}
where $c(M;\Sigma,\overline{\Sigma'})\in\C\setminus\{0\}$ is called the \emph{gluing anomaly factor} and depends only on the geometric data.
\end{itemize}

We proceed to make some more detailed comments on the axioms. A small technical ref\/inement compared to \cite{Oe:GBQFT,Oe:2dqym} is that the amplitude map in axiom (T4) is only required to be def\/ined on a dense subspace. This is sensible, since the amplitude map is generically not continuous. The def\/inition only on a dense subspace was to some extend implicit previously, but is now made completely explicit. This also af\/fects some of the other axioms.

\looseness=1
A more important change as compared to \cite{Oe:GBQFT,Oe:2dqym} is the splitting of what previously was the axiom (T5), formulating the gluing of two disjoint regions, into two parts, (T5a) and (T5b). This serves several purposes. Firstly, we may not only glue two disjoint regions together, but we may also perform gluings of dif\/ferent parts of the boundary of a connected region. Indeed, this was already actually used in \cite{Oe:2dqym}. On the other hand, any gluing of disjoint unions can of course be described by f\/irst joining the regions into a single (non-connected) region and then performing gluings on this region only. So there is no loss of generality involved in the change.

The most important change as compared to \cite{Oe:GBQFT,Oe:2dqym} is that we slightly weaken the gluing axiom. Namely, we introduce a scalar factor $c(M;\Sigma,\overline{\Sigma'})$ into the gluing equation~(\ref{eq:glueax1}). We call this the \emph{gluing anomaly factor}. It is a variant of what Turaev calls an \emph{anomaly cocycle}~\cite{Tur:qinv}, where the group in question is here the multiplicative group of complex numbers without~$0$. This yields another reason for the splitting of the gluing axiom into~(T5a) and~(T5b): The axiom~(T5a) does not involve such a factor.

The reasons for introducing the gluing anomaly factor are also several. The f\/irst and most obvious one is that in the quantization scheme we are going to present, axiom (T5b) would simply not be valid without this factor. Of course, this could be f\/ixed be introducing additional assumptions, but their general signif\/icance is not transparent at present. Another reason comes from the vacuum axioms (to be discussed in more detail below). Taken strictly, these would demand that the amplitude of a region without boundary is equal to~$1$.\footnote{The boundary Hilbert space in this case is $\C$ and the vacuum state there must correspond to the element $1\in\C$. Note also that axiom (T5a) combined with the fact that disjoint unions of regions are admissible means that we cannot make an ``exception'' for regions without boundary.} On the other hand, in simple examples of TQFTs with f\/inite-dimensional vector spaces the amplitude for a torus is equal to the dimension of the vector space associated with a single boundary component of a cylinder. This can be accommodated by ``moving'' this quantity into the anomaly factor. So we can make such examples of TQFTs compatible with both the core axioms and the vacuum axioms by utilizing the gluing anomaly factor. As a further remark note that this suggests that we formulate an explicit axiom that demands that the amplitude for a region without boundary be~$1$. However, since we do not want to even assume the existence of admissible regions without boundary, we do not do this. Moreover, as already mentioned, the vacuum axioms (when enforced) take care of this.

Finally, we mention that we have dropped the unitarity axiom~(T4b) as it does not seem to be satisf\/ied in the present quantization scheme. This does not mean, however, that we do not obtain unitary evolution when it is to be expected, see in particular Sections~\ref{sec:evol} and~\ref{sec:kg}. On the other hand, the precise physical signif\/icance of axiom (T4b) as formulated in~\cite{Oe:GBQFT,Oe:2dqym} is not transparent at present.

\subsection{Vacuum axioms}
\label{sec:vacax}

We list in the following the vacuum axioms, originally proposed in \cite{Oe:GBQFT}. The present version (including the numbering) is identical to that in \cite{Oe:2dqym}, except for the fact that we are in a setting without corners.
\begin{itemize}\itemsep=0pt
\item[(V1)] For each hypersurface $\Sigma$ there is a distinguished state
  $\psi_{\Sigma,0}\in\cH_\Sigma$, called the \emph{vacuum state}.
\item[(V2)] The vacuum state is compatible with the involution. That is,
  for any hypersurface $\Sigma$,
  $\psi_{\overline{\Sigma},0}=\iota_\Sigma(\psi_{\Sigma,0})$.
\item[(V3)] The vacuum state is compatible with
  decompositions. Suppose the hypersurface
  $\Sigma$ decomposes into components
  $\Sigma_1\cup\dots\cup\Sigma_n$. Then
  $\psi_{\Sigma,0}=
 \tau_{\Sigma_1,\dots,\Sigma_n;\Sigma}(\psi_{\Sigma_1,0}\tens\cdots\tens\psi_{\Sigma_n,0})$.
\item[(V5)] The amplitude of the vacuum state is
  unity. That is, for any region $M$, $\rho_M(\psi_{\partial M,0})=1$.
\end{itemize}

As remarked in \cite{Oe:GBQFT,Oe:2dqym} we can immediately verify that these axioms satisfy key properties that are more conventionally associated with a vacuum. Firstly, a vacuum state is normalized. This follows from combining axioms (V2), (V3), (V5) and (T3x). Secondly, the vacuum is stable under evolution.

\begin{prop}
\label{prop:unitevvac}
Let $M$ be a region such that its boundary decomposes into a disjoint union $\partial M=\Sigma\cup\overline{\Sigma'}$. Assume moreover that there is a unitary operator $U:\cH_\Sigma\to\cH_{\Sigma'}$ such that{\samepage
\begin{gather*}
\rho_M\circ\tau_{\Sigma,\overline{\Sigma'};\partial M}(\psi\tens\iota_{\Sigma'}(\psi'))=\langle \psi', U\psi\rangle_{\Sigma'}
\qquad\forall\, \psi\in\cH_\Sigma, \psi'\in\cH_{\Sigma'} .
\end{gather*}
Then, $U\psi_{\Sigma,0}=\psi_{\Sigma',0}$.}
\end{prop}
\begin{proof}
Since $\psi_{\Sigma,0}$ is normalized, so is $U\psi_{\Sigma,0}$. But its inner product with the also normalized state $\psi_{\Sigma',0}$ is $1$. So it must be identical to $\psi_{\Sigma',0}$.
\end{proof}

Conversely, suppose that for each pair $(\Sigma,\Sigma')$ of connected admissible hypersurfaces (with suitable orientations) there is a region $M$ such that $\partial M=\Sigma\cup\overline{\Sigma'}$. Moreover, assume that the amplitude for such a region takes the form of unitary evolution as in Proposition~\ref{prop:unitevvac}. Then, given an association of a normalized state to each hypersurface such that these states are related by evolution, these states will satisfy the vacuum axioms stated above. In fact, in this special case the choice of a vacuum in the sense of these axioms is equivalent to the choice of one normalized state on one hypersurface.

\section{Ingredients}
\label{sec:ingredients}

In the following we describe certain motivational and mathematical ingredients of the quantization scheme to be considered subsequently.

\subsection{Motivation from classical f\/ield theory}
\label{sec:classft}

In this section we consider certain aspects of Lagrangian f\/ield theory. While the facts we present are well known (see e.g.\ \cite{Woo:geomquant}), we recall them here from a particular perspective that provides a~key motivation for the further development in the subsequent parts of this paper.

Consider the following simple setting of a classical f\/ield theory. We suppose that the theory is def\/ined on a smooth spacetime manifold $T$ of dimension $d$ and determined by a f\/irst-order Lagrangian density $\Lambda(\varphi,\partial\varphi,x)$ with values in $d$-forms on~$T$. Here $x\in T$ denotes a point in spacetime, $\varphi$ a f\/ield conf\/iguration at a point and~$\partial\varphi$ the spacetime derivative at a point of a f\/ield conf\/iguration. We shall assume that the conf\/igurations are sections of a trivial vector bundle over $T$. We shall also assume in the following that all f\/ields decay suf\/f\/iciently rapidly at inf\/inity where required (i.e., where regions or hypersurfaces are non-compact).

Given a spacetime region $M$ and a f\/ield conf\/iguration $\phi$ in $M$ its action is given by
\begin{gather*}
 S(\phi)=\int_M \Lambda(\phi(\cdot),\partial\phi(\cdot),\cdot) .
\end{gather*}
Denote the space of f\/ield conf\/igurations in $M$ by $K_M$. The exterior derivative of the action yields a one-form $\xd S$ on $K_M$. Let $X\in K_M$ but think of it as an element of the tangent space~$T_\phi K_M$. The usual variational calculation (replace $\phi$ by $\phi+t X$ and keep only the f\/irst order in~$t$) yields
\begin{gather}
 \xd S_\phi(X)=\int_{\partial M} X^a \left.\partial_\mu\lrcorner\frac{\delta \Lambda}{\delta\, \partial_\mu\varphi^a}\right|_\phi
 + \int_M X^a \left.\left(\frac{\delta \Lambda}{\delta \varphi^a}-\partial_\mu \frac{\delta\Lambda}{\delta\, \partial_\mu\varphi^a}\right)\right|_\phi .
\label{eq:ds}
\end{gather}
Equating the term in the brackets to zero yields the Euler--Lagrange equations (in their version as $d$-forms). Supposing that $X$ vanishes on the boundary $\partial M$, $\xd S_\phi(X)=0$ if and only if $\phi$ satisf\/ies these equations. This is the usual variational principle.

We can also view $\xd S$ as a $1$-form on the space~$L_M$ of solutions of the Euler--Lagrange equations in~$M$, in which case the bulk term in (\ref{eq:ds}) vanishes. For this interpretation we also have to choose $X$ to be a solution of the linearized Euler--Lagrange equations around $\phi$, i.e., $X\in T_\phi L_M$. Since we are dealing with a pure boundary term this motivates the def\/inition of the $1$-form
\begin{gather*}
 \theta_\phi(X)\defeq\int_\Sigma X^a \left.\partial_\mu\lrcorner\frac{\delta \Lambda}{\delta\, \partial_\mu\varphi^a}\right|_\phi .
\end{gather*}
This can be def\/ined for an arbitrary hypersurface $\Sigma$. Moreover, it is naturally def\/ined on the space of solutions of the Euler--Lagrange equations in a neighborhood of~$\Sigma$, which we will denote by $L_\Sigma$. This $1$-form is usually called the \emph{symplectic potential}.

Taking the exterior derivative of the symplectic potential (viewed as a $1$-form on $L_\Sigma$) yields the usual \emph{symplectic form} on $L_\Sigma$:
\begin{gather}
\omega_\phi(X,Y)  =\xd\theta_\phi(X,Y)
 =\frac{1}{2}\int_\Sigma \left(\left. (X^b Y^a-Y^b X^a)\, \partial_\mu\lrcorner
 \frac{\delta^2\Lambda}{\delta\varphi^b\delta\,\partial_\mu\varphi^a}\right|_\phi\right. \nonumber\\
 \left.
 \hphantom{\omega_\phi(X,Y)  =\xd\theta_\phi(X,Y)=}{}
  +(Y^a\partial_\nu X^b-X^a \partial_\nu Y^b)\left.\partial_\mu\lrcorner
 \frac{\delta^2\Lambda}{\delta\,\partial_\nu\varphi^b\delta\,\partial_\mu\varphi^a}\right|_\phi\right) .
\label{eq:sympl}
\end{gather}
We shall assume that $\omega_\phi$ is non-degenerate. Note that $\omega$ changes its sign if the orientation of $\Sigma$ is reversed.

Consider again a region $M$ with boundary $\partial M$. We have a map $r_M:L_M\to L_{\partial M}$ by simply forgetting how the solutions look like in the interior of $M$. For any $\phi\in L_M$ this induces a map $(r_M^*)_\phi:T_\phi L_M\to T_{r_M(\phi)} L_{\partial M}$. It is then clear that for any $X,Y\in T_\phi L_M$ we have
\begin{gather*}
 \omega_{r_M(\phi)}((r_M^*)_\phi(X),(r_M^*)_\phi(Y))=\xd\xd S_\phi(X,Y)=0.
\end{gather*}
That is, $(r_M^*)_\phi(T_\phi L_M)$ is an \emph{isotropic subspace} of $T_{r_M(\phi)} L_{\partial M}$.

The above is most often used in the context where $T$ carries a Lorentzian metric and the Euler--Lagrange equations are hyperbolic partial dif\/ferential equations. The regions $M$ of interest are then the ones bounded by pairs $(\Sigma_1,\overline{\Sigma_2})$ of spacelike Cauchy hypersurfaces. Note that we can naturally identify the solution spaces $L_T=L_M=L_{\Sigma_1}=L_{\Sigma_2}$ by the Cauchy property. The symplectic form associated with $\partial M$ is then the sum of two components, $\omega_{\partial M}=\omega_{\Sigma_1}-\omega_{\Sigma_2}$, corresponding to the connected components of $\partial M$. Here we have taken the same orientation for $\Sigma_1$ and $\Sigma_2$, thus the minus sign coming from orientation reversal of $\Sigma_2$ with respect to the induced orientation as a boundary of $M$. One is then interested in global solutions, for which in particular $\omega_{\partial M}=0$ holds, i.e., $\omega_{\Sigma_1}=\omega_{\Sigma_2}$. Thus, one obtains a symplectic form independent of the choice of Cauchy hypersurface. This serves then as a good def\/inition of a symplectic form on the space of global solutions on $T$.

However, we do not want to specialize to this situation here, but rather insist in thinking of the symplectic form as def\/ined on the space of local solutions in a neighborhood of a hypersurface. Moreover, we want to use the abovementioned example of pairs of Cauchy hypersurfaces to motivate that the subspace $(r_M^*)_\phi(T_\phi L_M)\subseteq T_{r_M(\phi)} L_{\partial M}$ is in favorable cases not merely isotropic, but even \emph{Lagrangian}. Suppose the spaces of solutions in question were f\/inite-dimensional. Then we would have in the above example $\dim T_\phi L_{\partial M}=\dim T_\phi L_{\Sigma_1}+\dim T_\phi L_{\Sigma_2}=2\dim T_\phi L_M$. Thus, since $T_\phi L_M$ is isotropic and has half the dimension of $T_\phi L_{\partial M}$, it must be Lagrangian.

Above we have loosely talked about $L_\Sigma$ as the space of solutions of the Euler--Lagrange equations in a neighborhood of the hypersurface $\Sigma$. A better def\/inition would be to say that $L_\Sigma$ is the space of \emph{germs} of solutions near $\Sigma$. That is, we take the space of solutions in neighborhoods with two solutions being equivalent if they coincide in some sub-neighborhood of $\Sigma$. It is clear that for this to make sense one would usually require the hypersurfaces $\Sigma$ to carry some additional structure besides being submanifolds of codimension one.

\subsection{Elements of geometric quantization}
\label{sec:geomquant}

Continuing the discussion of the previous section we shall consider initial steps in a \emph{geometric quantization} of the spaces $L_\Sigma$ of local solutions on hypersurfaces $\Sigma$. More specif\/ically we shall only consider the case of a \emph{holomorphic} or \emph{K\"ahler} quantization. Furthermore, we shall restrict ourselves to the particularly simple case where the spaces $L_\Sigma$ are real vector spaces and can be naturally identif\/ied with all their tangent spaces. For much more details about geometric quantization, see \cite{Woo:geomquant}.

We suppose that for a given hypersurface $\Sigma$ we are given a real vector space $L_\Sigma$ of classical solutions near $\Sigma$. This comes equipped with a non-degenerate symplectic form $\omega_\Sigma$. The additional datum we need is a \emph{complex structure} $J_\Sigma:L_\Sigma\to L_\Sigma$ compatible with the symplectic structure. That is, $J_\Sigma$ is a linear map satisfying $J_\Sigma^2=-\id_\Sigma$ and $\omega_\Sigma(J_\Sigma(\cdot), J_\Sigma(\cdot))=\omega_\Sigma(\cdot,\cdot)$.
We then def\/ine the symmetric bilinear form $g_{\Sigma}:L_\Sigma\times L_\Sigma\to\R$ by
\begin{gather}
 g_\Sigma(\phi,\eta)\defeq 2\omega_\Sigma(\phi,J_\Sigma \eta) \qquad\forall\,\phi,\eta\in L_\Sigma .
\label{eq:kmetric}
\end{gather}
We shall assume that this form is positive def\/inite. The question of how $J_\Sigma$ or equivalently $g_\Sigma$ is obtained in practice is much less straightforward than that of the symplectic form $\omega_\Sigma$. See for example \cite{Wal:qftcurved} for a discussion of this point in the context of f\/ield theory in curved spacetime.

The next step is to complete $L_\Sigma$ to a real Hilbert space with the inner product $g_\Sigma$. (We will continue to write $L_\Sigma$ for this completion.) It is then true that the sesquilinear form
\begin{gather}
\{\phi,\eta\}_\Sigma\defeq g_\Sigma(\phi,\eta)+2\im\omega_\Sigma(\phi,\eta) \qquad\forall\, \phi,\eta\in L_\Sigma
\label{eq:kcip}
\end{gather}
makes $L_\Sigma$ into a complex Hilbert space, where multiplication with $\im$ is given by applying $J_\Sigma$.

The state space $\cH_\Sigma$ associated with $\Sigma$ is now constructed as follows: States are holomorphic functions on $L_\Sigma$ and form a Hilbert space with the inner product
\begin{gather}
 \langle \psi',\psi\rangle_\Sigma\defeq \int_{L_\Sigma} \psi(\phi)\overline{\psi'(\phi)}\exp\left(-\frac{1}{2} g_\Sigma(\phi,\phi)\right)\xd\mu(\phi).
\label{eq:gqip}
\end{gather}
Here $\mu$ is a ``translation-invariant measure'' on $L_\Sigma$. The exponential factor is what really comes from the more detailed theory of geometric quantization on which we shall not elaborate here. In fact, this formula works in the case were $L_\Sigma$ is f\/inite-dimensional. In the inf\/inite-dimensional case a measure $\mu$ of this kind does not exist. However, there is no dif\/f\/iculty in making the inf\/inite-dimensional case work properly as we shall see below. The space $\cH_\Sigma$ obtained in this way is nothing but the usual Fock space based on $L_\Sigma$ as the space of ``one-particle states'', but viewed as a space of holomorphic functions.

\subsection{Gaussian integration}
\label{sec:gint}

In the following we shall consider the Fock space $\cH$ generated by a separable Hilbert space $L$ (or rather its dual $V$). In particular, we shall be interested in viewing $\cH$ as a space of holomorphic functions following Bargmann \cite{Bar:hilbanalytic,Bar:remanalytic}. To this end we construct a suitable measure on an extension of $L$, via a projective limit based on measures on f\/inite-dimensional vector spaces. We start with some general considerations on projective limits of measure spaces, specializing step by step to our case of interest. In particular, this will allow us to make rigorous heuristic expressions such as~(\ref{eq:gqip}).

Let $\{L_\alpha\}_{\alpha\in A}$ be a projective system of measure spaces with surjective projections denoted by $l_{\alpha,\beta}:L_\alpha\tos L_\beta$. That is, each $L_\alpha$ comes equipped with a $\sigma$-algebra $\cM_\alpha$ and a measure $\nu_\alpha$. Moreover, the projections $l_{\alpha,\beta}$ are measure inducing, i.e., if $X\in \cM_\beta$ is measurable, then $l_{\alpha,\beta}^{-1}(X)$ is measurable and $\nu_\beta(X)=\nu_\alpha(l_{\alpha,\beta}^{-1}(X))$. What we would like to consider now is the projective limit $(\hat{L},\cM,\nu)$ in the category of measure spaces. However, in general it seems unclear whether this limit exists and how to construct it. Consider thus the limit merely in the category of sets with algebras of subsets and f\/initely additive measures. The projective limit exist and consists of the triple $(\hat{L},\cN,\nu)$ with the following properties: $\hat{L}$ is the projective limit $\hat{L}=\varprojlim L_\bullet$ in the category of sets. $\cN$ is the set of subsets of $\hat{L}$ that arise as preimages of measurable sets under the induced projections $l_\alpha:\hat{L}\to L_\alpha$. This is easily seen to be an algebra. It is then also clear that $\nu$ is well def\/ined and unique on $\cN$ and is f\/initely additive. If $\nu$ is in fact $\sigma$-additive on $\cN$, then we can use Hahn's theorem (see e.g.~\cite{Lan:rfanalysis}) to extend $\nu$ to the $\sigma$-algebra~$\cM$ generated by~$\cN$. Then $(\hat{L},\cM,\nu)$ would be our desired projective limit. We shall assume this for now, but will see explicitly below that this is satisf\/ied for the spaces and measures of particular interest to us. We let the $\sigma$-algebras $\cM_\alpha$ on $L_\alpha$ be completed with respect to the measures $\nu_\alpha$ and denote by~$\cM^*$ the completion of~$\cM$ with respect to the measure $\nu$.

Dually, denote by $\cM(L_\alpha)$ the vector space of complex valued functions on $L_\alpha$ that are $\cM_\alpha$-measurable. Then, the projections $l_{\alpha,\beta}$ induce injections $m_{\beta,\alpha}:\cM(L_\beta)\toi\cM(L_\alpha)$. Denoting measurable functions on $(\hat{L},\cM)$ by $\cM(\hat{L})$ we have the injections $m_\alpha:\cM(L_\alpha)\toi\cM(\hat{L})$ induced by the projections $l_\alpha$. We can think of these injections as inclusions and write for the injective limit $\varinjlim\cM(L_\bullet)\subseteq\cM(\hat{L})$. Similarly, for $1\le p\le \infty$ we consider the function spaces $\cL^p(L_\alpha,\nu_\alpha)$ forming an injective system. Then, for the injective limit we have $\varinjlim\cL^p(L_\bullet,\nu_\bullet)\subseteq\cL^p(\hat{L},\nu)$.

\begin{prop}
\label{prop:llpdense}
For $1\le p<\infty$ the subspace $\varinjlim\cL^p(L_\bullet,\nu_\bullet)$ is dense in $\cL^p(\hat{L},\nu)$.
\end{prop}
\begin{proof}
We know that the integrable simple functions are dense in $\cL^p(\hat{L},\nu)$. Furthermore, an integrable simple functions is a f\/inite linear combination of characteristic functions for sets of f\/inite measure. It is thus suf\/f\/icient to show that a characteristic function for a set of f\/inite measure $U\in\cM^*$ can be arbitrarily approximated by elements of $\varinjlim\cL^p(L_\bullet,\nu_\bullet)$. Let $\epsilon>0$. Using Hahn's Theorem we know that there is an element $V\in\cN$ such that $\nu((U\cup V)\setminus(U\cap V))<\epsilon$. But then, $\chi_V\in \varinjlim\cL^p(L_\bullet,\nu_\bullet)$ and $\|\chi_U-\chi_V\|_p<\epsilon^{1/p}$.
\end{proof}

We proceed to specialize to the case where the spaces $L_\alpha$ are f\/inite-dimensional real or complex vector spaces equipped with their standard topology. (We use in the following the notation $\K$ to denote a f\/ield that may either be $\R$ or $\C$.) More specif\/ically, we start with a~real or complex vector space $V$ and let $\{V_\alpha\}_{\alpha\in A}$ be the set of f\/inite-dimensional subspaces of~$V$. This set forms an injective system with injections $v_{\beta,\alpha}:V_\beta\toi V_\alpha$ and $v_\alpha:V_\alpha\toi V$ given by inclusions. We then def\/ine $L_\alpha$ to be the dual vector space of $V_\alpha$ yielding a projective system as already described. It is easy to see that $\hat{L}$ is the \emph{algebraic} dual of $V$. Equip the spaces $L_\alpha$ with the standard topology of f\/inite-dimensional vector spaces. Then, $\hat{L}$ can be seen as the projective limit in the category of topological spaces and as such carries the initial topology. Equivalently, this is the weak$^\star$ topology or topology of pointwise convergence as a space of linear functions on~$V$. In particular, $\hat{L}$ is in this way a complete locally convex Hausdorf\/f topological vector space. Moreover, we assume compatible measures $\nu_\alpha$ on the spaces $L_\alpha$ as described above which are inner regular Borel measures and let the associated $\sigma$-algebras $\cM_\alpha$ be completed with respect to the measures. In this setting $\nu$ is $\sigma$-additive on $\cN$ as is shown in the following using a result of Bochner, so that $(\hat{L},\cM,\nu)$ is a measure space.

\begin{dfn}
A projective system $\{Y_\beta\}_{\beta\in B}$ with projections $p_{\beta,\beta'}$ satisf\/ies the \emph{sequential maximality condition} if\/f for every sequence $\{\beta_k\}_{k\in\N}$ with $\beta_{k+1} > \beta_k$ and every sequence $\{y_k\}_{k\in\N}$ with $y_k\in Y_{\beta_k}$ such that $y_k=p_{\beta_{k+1},\beta_k}(y_{k+1})$, there is an element $y\in Y$ such that $y_k=p_{\beta_k}(y)$.
\end{dfn}

\begin{lem}
The system $\{L_\alpha\}_{\alpha\in A}$ satisfies the sequential maximality condition.
\end{lem}
\begin{proof}
Let $\{\alpha_k\}_{k\in\N}$ with $\alpha_{k+1}>\alpha_k$ and let $\{x_k\}_{k\in\N}$ with $x_k\in L_{\alpha_k}$ be such that $x_k=l_{\alpha_{k+1},\alpha_k}(x_{k+1})$. Consider the dual sequence $\{V_{\alpha_k}\}_{k\in\N}$ and set $W\defeq\bigcup_{k\in\N} V_{\alpha_k}$. The sequence $\{x_k\}_{k\in\N}$ def\/ines a linear map $x:W\to\K$ as follows. Let $a\in W$. Then, there exists a smallest $n\in\N$ such that $a\in V_{\alpha_n}$. Set $x(a)\defeq (x_n,a)_{\alpha_n}$ and note that $x(a)=(x_k,a)_{\alpha_k}$ for all $k\ge n$. It is easy to see that this prescription does indeed def\/ine a linear map $W\to\K$. Equipping $V$ for example with the weak$^\star$ topology with respect to $\hat{L}$, we can apply the Hahn--Banach theorem to extend $x$ to a linear map $V\to\K$. Thus, $x\in L$ and $l_{\alpha_k}(x)=x_k$, as required.
\end{proof}

\begin{thm}[Bochner {\cite[Theorem~5.1.1]{Boc:analysisprob}}]
\label{thm:bochner}
Let $\{L_\alpha\}_{\alpha\in A}$ be a projective system of Hausdorff spaces with inner regular Borel measures. Assume furthermore, that the system satisfies the sequential maximality condition. Then, the induced function $\nu:\cN\to [0,\infty]$ is $\sigma$-additive.
\end{thm}

Adding another layer of structure we suppose that $V$ is a real or complex separable Hilbert space. Any subspace $V_\alpha$ inherits this structure and the injections $v_{\beta,\alpha}:V_\beta\toi V_\alpha$ as well as $v_\alpha:V_\alpha\toi V$ become linear isometries. Moreover, by duality the spaces $L_\alpha$ also become Hilbert spaces in this way and the projections $l_{\alpha,\beta}:L_\alpha\tos L_\beta$ become partial linear isometries. Let $L$ be the topological dual of $V$ and thus also a Hilbert space. It is then clear that there is a natural inclusion $i:L\toi\hat{L}$ which is continuous.

We use the inner products on the spaces $L_\alpha$ to def\/ine Gaussian measures $\nu_\alpha$. In the case that the vector spaces are complex we forget their complex structure for the moment and regard them as real vector spaces. Note that a complex Hilbert space is canonically a real Hilbert space by taking the real part of the inner product. We f\/irst recall some elementary facts about Gaussian measures on f\/inite-dimensional vector spaces. Let $Q$ be a real positive def\/inite symmetric $n\times n$-matrix. Let $\mu$ be the Lebesgue measure on $\R^n$ and $\nu_Q$ the Gaussian measure given by
\begin{gather*}
 \xd\nu_Q(x)\defeq\exp\left(-\sum_{i,j} x_i Q_{i,j} x_j\right)
  \sqrt{\frac{\det Q}{\pi^n}} \xd\mu(x) .
\end{gather*}
Then, $\nu_Q$ is an inner and outer regular Borel probability measure. For later use we also remark the following lemmas.
\begin{lem}
Let $\one$ denote the $n\times n$ unit matrix and $r>0$. Then,
\begin{gather}
 \nu_\one(B_r(0))=\begin{cases}
 \displaystyle  e^{-r^2}\sum\limits_{k=n/2}^{\infty} \frac{1}{k!} r^{2 k} & \text{if $n$ is even},\\
  \displaystyle e^{-r^2}\sum\limits_{k=(n+1)/2}^{\infty}\frac{1}{\Gamma(k+1/2)} r^{2 k -1}
  & \text{if $n$ is odd} .
  \end{cases}
 \label{eq:rnvolball}
\end{gather}
\end{lem}

\begin{lem}
\label{lem:wick}
Suppose $f$ is a polynomial on $\R^n$ with complex coefficients,
\begin{gather*}
 f(x)=\sum_k f_k x_1^{k_1}\cdots x_n^{k_n} ,
\end{gather*}
where $k$ is a multi-index $k=(k_1,\dots,k_n)$.
Then, we have a version of Wick's theorem:
\begin{gather*}
 \int_{\R^n} f(x)\,\xd\nu_Q(x)=
 \sum_{m=0}^\infty \frac{1}{m! 4^m}\sum_{j_1,\dots,j_m}\sum_{l_1,\dots,l_m}
  ([l]+ [j])!  \big(Q^{-1}\big)_{l_1,j_1}\cdots\big(Q^{-1}\big)_{l_m,j_m}  f_{[l]+[j]} .
\end{gather*}
Here, the sums run over $j_1,\dots,j_m,l_1,\dots,l_m\in\{1,\dots,n\}$. The expressions $[j]$ and $[l]$ are multi-indices defined as follows. $[j]=([j]_1,\dots,[j]_n)$ where $[j]_i$ is the number of the indices $j_1,\dots,j_m$ that take the value~$i$. That is, $[j]_i\defeq |\{a\in\{1,\dots,m\}|j_a=i\}|$. Also $k!\defeq k_1!\cdots k_n!$ for a~multi-index.
\end{lem}

We proceed to assign measures $\nu_\alpha$ to the spaces $L_\alpha$. Given a basis $\{\eta_1,\dots,\eta_n\}$ of $L_\alpha$ as a~real Hilbert space def\/ine the positive def\/inite symmetric matrix $(Q_\alpha)_{i,j}\defeq (\eta_i,\eta_j)_{\R,L_\alpha}$ and the vector space isomorphism $\R^n\to L_\alpha$ given by $\eta:(x_1,\dots,x_n)\mapsto \sum\limits_{i=1}^n x_i\eta_i$. Now set $\nu_\alpha(B)\defeq \nu_{Q_\alpha}(\eta^{-1}(B))$ for $B$ a~Borel set in $L_\alpha$. Notice that if $\{\xi_1,\dots,\xi_n\}$ is the dual basis of $V_\alpha$, then $(Q_\alpha^{-1})_{i,j}=(\xi_i,\xi_j)_{\R,V_\alpha}$.

We have to show that this def\/inition is compatible with the projections $l_{\alpha,\beta}$. Thus, let $\alpha,\beta\in A$ such that $V_\beta\subseteq V_\alpha$ and choose a basis $\{\xi_1,\dots,\xi_n\}$ of $V_\alpha$ such that $\{\xi_1,\dots,\xi_k\}$ is a~basis of $V_\beta$. We notice that $(Q_\alpha^{-1})_{i,j}=(Q_\beta^{-1})_{i,j}$ for all $i,j\in\{1,\dots,k\}$. Let $f$ be a polynomial on~$L_\beta$. Then, with Lemma~\ref{lem:wick} we see that $\int_{L_\alpha} f\circ l_{\alpha,\beta}\,\xd\nu_\alpha =\int_{L_\beta} f\,\xd\nu_\beta$. Since polynomials are dense in  $\cL^1(\R^k,\nu_{Q_\beta})$, the same holds if we replace $f$ with any characteristic function of a~Borel set, implying the compatibility of the measures.

In case the vector spaces are really complex vector spaces we can alternatively use the complex inner product to def\/ine positive def\/inite Hermitian matrices $(Q_\alpha)_{i,j}\defeq (\eta_i,\eta_j)_{L_\alpha}$. A positive def\/inite Hermitian matrix def\/ines a measure on $\C^n$ via
\begin{gather}
 \xd\nu_Q(x)\defeq\exp\left(-\sum_{i,j} \overline{x_i} Q_{i,j} x_j\right)
  \frac{\det Q}{\pi^n} \xd\mu(x) .
\label{eq:mescq}
\end{gather}
However, these measures are \emph{identical} to those obtained by forgetting the complex structure as described above. This is easily seen to be due to the fact that the measures only depend on the quadratic form induced by the inner product which is the same in both cases.

Note that $L$ is dense in $\hat{L}$, but has measure zero, as the following Proposition shows. In particular, we cannot simply forget about $\hat{L}$ and restrict to $L$. This is not surprising since it is well known that there does not exist a Gaussian measure on an inf\/inite-dimensional Hilbert space~\cite{Sud:qintmes}.

\begin{prop}
Let $V$ be an infinite-dimensional separable Hilbert space. Then, $L\in\cM^*$ and $\nu(L)=0$.
\end{prop}

\begin{proof}
Given an ON-basis $\{\xi_i\}_{i\in\N}$ of $V$ (as a real or complex Hilbert space) we consider the sequence $\{\alpha_n\}_{n\in\N}$ where $\alpha_n\in A$ such that $V_{\alpha_n}$ is generated by $\{\xi_1,\dots,\xi_n\}$. Now f\/ix $r>0$ and let $B_{r,n}$ be the open ball of radius $r$ around $0$ in the Hilbert space $L_{\alpha_n}$. Denote the preimage $l_{\alpha_n}^{-1}(B_{r,n})$ in $L$ also by $B_{r,n}$. Observe that for $m\ge n$ we have $B_{r,m}\subset B_{r,n}$ in $L$. Thus, we obtain a decreasing sequence $\{B_{r,n}\}_{n\in\N}$ of measurable subsets of $L$. Their intersection $B_{r}'\defeq\bigcap_{n\in\N} B_{r,n}$ is thus also a measurable set in $L$ and we have
\begin{gather*}
 \nu(B_{r}')=\lim_{n\to\infty} \nu(B_{r,n}) .
\end{gather*}
However, recalling formula~(\ref{eq:rnvolball}) we see that the volumes of the balls tend to zero as $n$ tends to inf\/inity, so $\nu(B_r')=0$. (In the complex case~$n$ in formula~(\ref{eq:rnvolball}) is to be replaced with~$2n$.) Now def\/ine $B_r\subset L$ as the open ball with radius $r$ in the Hilbert space~$L$. Then, $B_r\subseteq B_r'$. In particular, $B_r$ is measurable (with respect to $\cM^*$) and $\nu(B_r)=0$. Now consider the countable union $\bigcup_{l\in\N} B_l$ of balls with integer radius, f\/illing out $L$ completely. Since each has measure zero, the measure of their union, that is~$L$, is zero as well.
\end{proof}

Note that elements of $\varinjlim \cM(L_\bullet)$ can be uniquely represented as functions on $L$. To see this consider the following Lemma.

\begin{lem}
\label{lem:injl}
For any $\alpha\in A$ there is a unique isometric injection $i_\alpha:L_\alpha\toi L$ such that $l_\alpha\circ i\circ i_\alpha=\id_{L_\alpha}$.
\end{lem}
\begin{proof}
Given $\alpha\in A$ let $C$ be the subspace of $L$ such that $(c,v)=0$ for all $c\in C$ and $v\in V_\alpha$. It is now easy to see that $L_\alpha$ can be identif\/ied with $C^\perp$ as spaces of linear functions on $V_\alpha$. The identity $l_\alpha\circ i\circ i_\alpha=\id_{L_\alpha}$ is then also clear as well as the fact that $i_\alpha$ is isometric.
\end{proof}

Recall that any element of $\varinjlim \cM(L_\bullet)$ arises as a (measurable) function $f:L_\alpha\to\C$. Then, the induced function $f\circ l_\alpha\circ i:L\to\C$ uniquely determines $f$ by the above Lemma. Applying this to $\varinjlim\cL^p(L_\bullet,\nu_\bullet)$ suggests together with Proposition~\ref{prop:llpdense} that for certain $\cL^p$-functions that ``behave nicely'' under approximation, their restriction to $L$ is already suf\/f\/icient to determine them completely. We shall see in the next section that this is indeed true for holomorphic $\cL^2$-functions.

\begin{dfn}
Let $f:L\to\C$. If there exists a closed subspace $C\subseteq L$ of f\/inite codimension such that $f(x+c)=f(x)$ for all $x\in L$ and $c\in C$, then we call $f$ \emph{almost translation invariant}.
\end{dfn}

\begin{prop}
\label{prop:evalgint}
Let $f:L\to\C$ be almost translation invariant with respect to the closed subspace $C\subseteq L$ of finite codimension. Set $\alpha\in A$ such that $C=\ker (l_\alpha\circ i)$ and define $\hat{f}:\hat{L}\to\C$ by $\hat{f}\defeq f\circ i_\alpha\circ l_\alpha$. Let $\{\eta_1,\dots,\eta_n\}$ be a basis of $C^\perp\subseteq L$ and define the $n\times n$ matrix $Q_{i,j}\defeq (\eta_i,\eta_j)_{L}$. Furthermore define $\tilde{f}:\K^n\to\C$ by $\tilde{f}(x)\defeq f\big(\sum\limits_{i=1}^n x_i\eta_i\big)$. If for $p\in [1,\infty]$ we have $\tilde{f}\in\cL^p(\K^n,\nu_Q)$ then we also have $\hat{f}\in\cL^p(\hat{L},\nu)$ and $\|\tilde{f}\|_p=\|f\|_p$. Moreover, if $\tilde{f}\in\cL^1(\K^n,\nu_Q)$ we have
\begin{gather*}
 \int_{\hat{L}} \hat{f}\,\xd\nu = \int_{\K^n} \tilde{f}\,\xd\nu_Q .
\end{gather*}
\end{prop}
\begin{proof}
This follows from our above prescription for the measures $\nu_\alpha$ combined with Lemma~\ref{lem:injl} and the above discussion.
\end{proof}

Note that in the above proposition we have not explicitly mentioned whether $L$ is a real or a~complex Hilbert space. Indeed, if $L$ is a complex Hilbert space the Proposition is valid \emph{both} for~$L$ as a complex Hilbert space and as a real Hilbert space with the induced inner product.

\begin{prop}
\label{prop:rtransinv}
Let $x\in L$ and $1\le p<\infty$. Then the map $f\mapsto f_x$, where
\begin{gather*}
 f_x(z)\defeq f(z+x) \exp\left(-\frac{1}{p}\Re(2z + x, x)_L\right)
\end{gather*}
is an isometric isomorphism $\cL^p(\hat{L},\nu)\to\cL^p(\hat{L},\nu)$. Moreover, in the case $p=1$ this preserves the integral and in the case $p=2$ the inner product.
\end{prop}
\begin{proof}
For the dense subspace $\varinjlim\cL^p(L_\bullet,\nu_\bullet)$ this follows using translation invariance of the Lebesgue measure. Given a Cauchy sequence $\{f_n\}_{n\in\N}$ in $\varinjlim\cL^p(L_\bullet,\nu_\bullet)$ that converges to some $f\in\cL^p(\hat{L},\nu)$ we observe that it has a subsequence that converges pointwise almost everywhere to~$f$. Then, the corresponding subsequence of $\{(f_n)_x\}_{n\in\N}$ converges pointwise almost everywhere to~$f_x$. Hence $\|f\|_p=\|f_x\|_p$ and $\int f=\int f_x$ if $p=1$.
\end{proof}

\subsection{Holomorphic functions and coherent states}
\label{sec:hcoh}

Assume in this section that $L$ is a complex Hilbert space. The space $\cL^2(\hat{L},\nu)$ carries the usual (non-def\/inite) inner product
\begin{gather}
 \langle f,g\rangle\defeq \int_{\hat{L}} \overline{f(x)} g(x)\,\xd\nu(x) ,
\label{eq:hip}
\end{gather}
making it complete. We shall be interested particularly in the subspace of this space of \emph{holomorphic} functions. This will make precise the formula~(\ref{eq:gqip}) and its setting.

Consider f\/irst the f\/inite-dimensional setting of $\C^n$ with a positive def\/inite inner product $(\cdot,\cdot)_Q$ given by a positive def\/inite Hermitian $n\times n$ matrix $Q$ via $(x,y)_Q\defeq \sum_{i,j}\overline{x_i}Q_{i,j} y_j$. Equip~$\C^n$ with the probability measure $\nu_Q$ as described in (\ref{eq:mescq}). Let $\rH^2(\C^n,\nu_Q)\subseteq\cL^2(\C^n,\nu_Q)$ denote the Hilbert space of square integrable holomorphic functions on $\C^n$ with respect to $\nu_Q$. This is a~reproducing kernel Hilbert space, i.e., it is a~Hilbert space of functions on which the evaluation $f\mapsto f(x)$ is continuous. We recall some well known facts about this space~\cite{Bar:hilbanalytic}. By the Riesz representation theorem, there is for each $x\in\C^n$ a unique element $K_x$ of the Hilbert space that realizes this evaluation. That is
\begin{gather}
\langle K_x,f\rangle=f(x)\qquad \forall\, x\in\C^n, \quad \forall \, f\in\rH^2\big(\C^n,\nu_Q\big).
\label{eq:evalcn}
\end{gather}
Indeed, as is easily verif\/ied,
\begin{gather*}
 K_x(z)=\exp\left((x,z)_Q\right) \qquad\forall\,  x,z\in\C^n .
\end{gather*}
In particular, we have
\begin{gather*}
 \langle K_x, K_y\rangle = K_y(x)=\exp\left((y,x)_Q\right)\qquad\forall\, x,y\in\C^n .
\end{gather*}
We also have the completeness relation
\begin{gather*}
 \langle f,g\rangle=\int \langle f, K_z\rangle \langle K_z, g\rangle\, \xd\nu_Q(z)\qquad\forall\, f,g\in\rH^2\big(\C^n,\nu_Q\big) .
\end{gather*}

\begin{prop}
\label{prop:cohdense}
The vector space of finite linear combinations of evaluations $\{K_x\}_{x\in\C^n}$ is dense in $\rH^2(\C^n,\nu_Q)$.
\end{prop}
\begin{proof}
Let $C$ be the closure of the subspace generated by $\{K_x\}_{x\in\C^n}$ and $C^\perp$ its orthogonal complement. Let $f\in C^\perp$. Then, $f(x)=\langle K_x,f\rangle=0$ for all $x\in\C^n$, i.e., $f=0$. Hence, $C^\perp=\{0\}$ and $C=\rH^2(\C^n,\nu_Q)$.
\end{proof}

For $\alpha\in A$ let $\rH^2(L_\alpha,\nu_\alpha)\subseteq\cL^2(L_\alpha,\nu_\alpha)$ denote the Hilbert space of square integrable holomorphic functions on $L_\alpha$. Then, $\varinjlim \rH^2(L_\bullet,\nu_\bullet)\subseteq\varinjlim \cL^2(L_\bullet,\nu_\bullet)$ is a (def\/inite) inner product space in the obvious way. Note that all elements of $\varinjlim \rH^2(L_\bullet,\nu_\bullet)$ have unique representations as functions on $L$ since they are almost translation invariant. We shall now be interested in the Hilbert space $\rH^2(\hat{L},\nu)$, def\/ined as the completion of $\varinjlim \rH^2(L_\bullet,\nu_\bullet)$. Any element of $\rH^2(\hat{L},\nu)$ can be represented concretely as an element of $\cL^2(\hat{L},\nu)$. Moreover, as we shall see, it has a unique representation as a holomorphic function on $L$.

\begin{dfn}
We say that a function $f:L\to\C$ is \emph{holomorphic} if\/f $f$ is continuous, bounded on every ball and holomorphic (in the usual sense) at every point in every direction. We denote the space of holomorphic functions on $L$ by $\rH(L)$. We equip it with the topology of uniform convergence on balls.
\end{dfn}
Note that $\rH(L)$ is a locally convex, metrizable and complete topological vector space.

Consider the family of functions $\{K_x\}_{x\in L}$ with $K_x:\hat{L}\to\C$ given by $z\mapsto \exp((x, z)_L)$ for all $z\in \hat{L}$. Here as in the following we shall use the notation $(\cdot,\cdot)_L$ to not only denote the complex inner product in $L$, but also its canonical extensions to sesquilinear maps $\hat{L}\times L\to\C$ and $L\times \hat{L}\to\C$. As we shall see these functions generalize the evaluation functions considered above and we will call them \emph{coherent states} in accordance with the language of quantum theory.

\begin{prop}
\label{prop:csinf}
The functions $K_x$ have the following properties:
\begin{itemize}\itemsep=0pt
\item[$1.$] $K_x\in\varinjlim\rH^2(L_\bullet,\nu_\bullet)$ for all $x\in L$.
\item[$2.$] $\langle K_x, f\rangle = f(x)$ for all $f\in \varinjlim\rH^2(L_\bullet,\nu_\bullet)$ and $x\in L$.
\item[$3.$] $\langle K_x, K_y\rangle=\exp\left((y,x)_{L}\right)$ for all $x,y\in L$.
\item[$4.$] $\|K_x\|_2=\exp\left(\frac{1}{2}\|x\|_2^2\right)$ for all $x\in L$.
\end{itemize}
\end{prop}
\begin{proof}
1.~$K_x$ restricted to $L$ is translation invariant with respect to the orthogonal complement in $L$ of the $1$-dimensional (or $0$-dimensional) subspace generated by~$x$.
2.~Given $f\in\varinjlim\rH^2(L_\bullet,\nu_\bullet)$ there exist $\alpha,\beta\in A$ such that $K_x\in\rH^2(L_\alpha,\nu_\alpha)$ and $f\in\rH^2(L_\beta,\nu_\beta)$. Moreover, there is a $\gamma\in A$ such that there are injections $l_{\gamma,\beta}:L_\gamma\to L_\beta$ and $l_{\gamma,\alpha}:L_\gamma\to L_\alpha$ with $\alpha$ as above. We can thus view $f$ and $K_x$ both as elements of $\rH^2(L_\gamma,\nu_\gamma)$. In particular $K_x$ is then an evaluation on a f\/inite-dimensional complex vector spaces as described above and we can apply~(\ref{eq:evalcn}).
3.~This arises as a simple consequence of 2.
4.~This is in turn a simple consequence of~3.
\end{proof}

\begin{prop}
\label{prop:cohdensei}
The vector space of finite linear combinations of coherent states is dense in $\rH^2(\hat{L},\nu)$.
\end{prop}
\begin{proof}\sloppy
Recall that every element of $\varinjlim\rH^2(L_\bullet,\nu_\bullet)$ can be represented as an element of $\rH^2(L_\alpha,\nu_\alpha)$ for some $\alpha$. Combining this with Proposition~\ref{prop:cohdense} yields denseness in $\varinjlim\rH^2(L_\bullet,\nu_\bullet)$. But since $\varinjlim\rH^2(L_\bullet,\nu_\bullet)$ is dense in $\rH^2(\hat{L},\nu)$ by def\/inition, the statement follows.
\end{proof}

\begin{lem}
\label{lem:injholoml}
Let $f\in\varinjlim\rH^2(L_\bullet,\nu_\bullet)$ and $r>0$. Then,
\begin{gather}
 \sup_{x\in \overline{B_r(0)}} |f(x)|\le\exp\left(\frac{1}{2}r^2\right)\|f\|_2.
\label{eq:sup2norm}
\end{gather}
In particular, the natural continuous linear map $\varinjlim\rH^2(L_\bullet,\nu_\bullet)\toi \rH(L)$ is injective.
\end{lem}
\begin{proof}
By the Cauchy--Schwarz inequality we have
\begin{gather*}
 |f(x)|=|\langle K_x, f\rangle|\le \|K_x\|_2 \|f\|_2\qquad
 \forall \, x\in L, \quad \forall \, f\in \varinjlim\rH^2(L_\bullet,\nu_\bullet)
\end{gather*}
But $\|K_x\|_2=\exp(\frac{1}{2}\|x\|_2^2)$ according to Proposition~\ref{prop:csinf}.4. This yields~(\ref{eq:sup2norm}). Notice that each element of $\varinjlim\rH^2(L_\bullet,\nu_\bullet)$ is continuous, bounded in any ball and holomorphic at every point in every direction. Hence there is an injective linear map $\varinjlim\rH^2(L_\bullet,\nu_\bullet)\toi \rH(L)$ in the obvious way. Its continuity follows precisely from~(\ref{eq:sup2norm}) since it shows that the composition of this map with every seminorm def\/ining the topology of $\rH(L)$ is continuous.
\end{proof}

\begin{lem}
\label{lem:hconvzero}
Let $\{f_k\}_{k\in\N}$ be a sequence in $\varinjlim\rH^2(L_\bullet,\nu_\bullet)$ converging to $f\in\rH^2(\hat{L},\nu)$. Suppose that $\{f_k(x)\}_{k\in\N}$ converges to zero for all $x\in L$. Then, $f=0$.
\end{lem}
\begin{proof}
For all $x\in L$ observe
\begin{gather*}
 \langle K_x, f\rangle = \lim_{k\to\infty} \langle K_x, f_k\rangle
 =\lim_{k\to\infty} f_k(x)=0 .
\end{gather*}
But recall that the coherent states are dense according to Proposition~\ref{prop:cohdensei}. So $f=0$.
\end{proof}

\begin{thm}
\label{thm:h2holom}
There is a natural injective continuous linear map $\rH^2(\hat{L},\nu)\toi \rH(L)$.
\end{thm}
\begin{proof}
The map is the one given by Lemma~\ref{lem:injholoml} on $\varinjlim\rH^2(L_\bullet,\nu_\bullet)$. Since $\rH(L)$ is complete, its extension to a linear map $\rH^2(\hat{L},\nu)\to \rH(L)$ is immediate. It remains to verify its injectivity. Let $f\in \rH^2(\hat{L},\nu)$ such that its image in $\rH(L)$ is zero. Consider a sequence $\{f_k\}_{k\in\N}$ of elements of $\varinjlim\rH^2(L_\bullet,\nu_\bullet)$ that converge to $f$. Then, the image of the sequence converges to zero in $\rH(L)$. In, particular $\{f_k(x)\}_{k\in\N}$ converges to zero for every $x\in L$. But by Lemma~\ref{lem:hconvzero} this implies that $f=0$.
\end{proof}

\begin{prop}
\label{prop:coheval}
Let $x\in L$. Then, $\langle K_x, f\rangle = f(x)$ for all $f\in \rH^2(\hat{L},\nu)$ viewed as elements of $\rH(L)$. In the same sense, the completeness relation holds
\begin{gather*}
 \langle f, g\rangle=\int_{\hat{L}} \langle f ,K_x\rangle
  \langle K_x, g\rangle\, \xd\nu(x) \qquad\forall\, f,g\in \rH^2\big(\hat{L},\nu\big) .
\end{gather*}
\end{prop}
\begin{proof}
Combine Proposition~\ref{prop:csinf} with Theorem~\ref{thm:h2holom}, the denseness of $\varinjlim\rH^2(L_\bullet,\nu_\bullet)$ in $\rH(L)$ and the fact that evaluation is continuous on $\rH(L)$. The completeness relation is then also obvious.
\end{proof}

This proposition means in particular that we may view $\rH^2(\hat{L},\nu)$ as a reproducing kernel Hilbert space of functions on~$L$.

\begin{prop}
The Hilbert space $\rH^2(\hat{L},\nu)$ is separable.
\end{prop}
\begin{proof}
Since $L$ is separable, there is a countable dense subset $D\subseteq L$. Since the subspace of linear combinations of coherent states is dense in $\rH^2(\hat{L},\nu)$ by Proposition~\ref{prop:cohdensei} it is suf\/f\/icient to show that any multiple $\lambda K_x$ of a coherent state can be arbitrarily approximated by a multiple of a~coherent state of the form $q K_p$, where $p\in D$ and $q\in\Q\oplus\im\Q$. This is easy to verify by explicit calculation.
\end{proof}

\section{Quantization}
\label{sec:quantization}

\subsection{Classical data}
\label{sec:classax}

We assume the classical theory is provided in the following form.
\begin{itemize}\itemsep=0pt
\item[(C1)] Associated to each hypersurface $\Sigma$ is a complex separable Hilbert space $L_\Sigma$ (thought of as the space of solutions near $\Sigma$) with inner product denoted by $\{\cdot ,\cdot\}_\Sigma$. We also def\/ine $g_\Sigma(\cdot,\cdot)\defeq \Re\{\cdot ,\cdot\}_\Sigma$ and $\omega_\Sigma(\cdot,\cdot)\defeq \frac{1}{2}\Im\{\cdot ,\cdot\}_\Sigma$ and denote by $J_\Sigma:L_\Sigma\to L_\Sigma$ the scalar multiplication with $\im$ in $L_\Sigma$.
\item[(C2)] Associated to each hypersurface $\Sigma$ there is a conjugate linear involution $L_\Sigma\to L_{\overline\Sigma}$ under which the inner product is complex conjugated. We will not write this map explicitly, but rather think of $L_\Sigma$ as identif\/ied with $L_{\overline\Sigma}$. Then, $\{\phi',\phi\}_{\overline{\Sigma}}=\overline{\{\phi',\phi\}_\Sigma}$ for all $\phi,\phi'\in L_\Sigma$.
\item[(C3)] Suppose the hypersurface $\Sigma$ decomposes into a disjoint
  union of hypersurfaces \mbox{$\Sigma=\Sigma_1\cup\cdots$} $\cdots\cup\Sigma_n$. Then,
  there is an isometric isomorphism of complex Hilbert spaces
  $L_{\Sigma_1}\oplus\cdots$ \mbox{$\cdots\oplus L_{\Sigma_n}\to L_\Sigma$}. We will not write this map explicitly, but rather think of it as an identif\/ication.
\item[(C4)] Associated to each region $M$ is a real vector space $L_M$ (thought of as the space of solutions in~$M$).

\item[(C5)] Associated to each region $M$ there is a linear map of real vector spaces $r_M:L_M\to L_{\partial M}$. The image $L_{\tilde{M}}$ of $r_M$ is a real closed subspace of $L_{\partial M}$. Furthermore it is Lagrangian with respect to the symplectic form $\omega_{\partial M}$.
\item[(C6)] Let $M_1$ and $M_2$ be regions and $M\defeq M_1\cup M_2$ be their disjoint union. Then $L_{M}=L_{M_1}\oplus L_{M_2}$. Moreover, $r_M=r_{M_1}+r_{M_2}$.
\item[(C7)] Let $M$ be a region with its boundary decomposing as a disjoint union $\partial M=\Sigma_1\cup\Sigma\cup \overline{\Sigma'}$, where $\Sigma'$ is a copy of $\Sigma$. Let $M_1$ denote the gluing of $M$ to itself along $\Sigma$, $\overline{\Sigma'}$ and suppose that $M_1$ is a region. Note $\partial M_1=\Sigma_1$. Then, there is an injective linear map $r_{M;\Sigma,\overline{\Sigma'}}:L_{M_1}\toi L_{M}$ such that
\begin{gather*}
 L_{M_1}\toi L_{M}\rightrightarrows L_\Sigma
\end{gather*}
is an exact sequence. Here the arrows on the right hand side are compositions of the map $r_M$ with the orthogonal projections of $L_{\partial M}$ to $L_\Sigma$ and $L_{\overline{\Sigma'}}$ respectively (the latter identif\/ied with $L_\Sigma$). Moreover, the following diagram commutes, where the bottom arrow is the orthogonal projection
\begin{gather*}
\begin{split}
\xymatrix{
  L_{M_1} \ar[rr]^{r_{M;\Sigma,\overline{\Sigma'}}} \ar[d]_{r_{M_1}} & & L_{M} \ar[d]^{r_{M}}\\
  L_{\partial M_1}  & & L_{\partial M} \ar[ll]}
  \end{split}
\end{gather*}
\end{itemize}

We add the following observations: $g_\Sigma$ is a real positive def\/inite symmetric bilinear form making $L_\Sigma$ into a real Hilbert space. $\omega_\Sigma$ is a real anti-symmetric non-degenerate bilinear form making $L_\Sigma$ into a symplectic vector space. Also
\begin{gather*}
g_{\overline{\Sigma}}=g_\Sigma,\qquad J_{\overline{\Sigma}}=-J_{\Sigma},\qquad \omega_{\overline{\Sigma}}=-\omega_{\Sigma} .
\end{gather*}
Moreover, for all $\phi,\phi'\in L_\Sigma$
\begin{alignat*}{3}
& g_\Sigma(\phi,\phi')   = 2\omega_\Sigma(\phi, J_\Sigma\phi'), \qquad &&
  \{\phi,\phi'\}_\Sigma   = g_\Sigma(\phi,\phi')
  +2\im\omega_\Sigma(\phi,\phi'), & \\
& g_\Sigma(\phi,\phi')   = g_\Sigma(J_\Sigma \phi, J_\Sigma\phi'), \qquad &&
  \omega_\Sigma(\phi,\phi')   =\omega_\Sigma(J_\Sigma\phi,J_\Sigma \phi') . &
\end{alignat*}

\begin{lem}
\label{lem:decis}
Let $M$ be a region. Then, $L_{\partial M}$ understood as a real Hilbert space decomposes into an orthogonal direct sum $L_{\partial M}=L_{\tilde{M}}\oplus J_{\partial M} L_{\tilde{M}}$.
\end{lem}
\begin{proof}
Let $\phi, \phi'\in L_{\tilde{M}}$. Then, $g_{\partial M}(\phi, J_{\partial M}\phi')=-2\omega_{\partial M}(\phi,\phi')=0$, since $L_{\tilde{M}}$ is isotropic in~$L_{\partial M}$. Thus, $L_{\tilde{M}}$ and $J_{\partial M} L_{\tilde{M}}$ are orthogonal in $L_{\partial M}$. Now suppose that $\phi\in L_{\partial M}$ is in the orthogonal complement of $L_{\tilde{M}}$. That is $g_{\partial M}(\phi',\phi)=2\omega_{\partial M}(\phi',J_{\partial M}\phi)=0$ for all $\phi'\in L_{\tilde{M}}$. But since $L_{\tilde{M}}$ is coisotropic in $L_{\partial M}$ we must have $J_{\partial M}\phi\in L_{\tilde{M}}$, i.e., $\phi\in J_{\partial M} L_{\tilde{M}}$. Hence, $L_{\partial M}=L_{\tilde{M}}+ J_{\partial M} L_{\tilde{M}}$.
\end{proof}

This means in particular, that we may view $L_{\partial M}$ as the complexif\/ication of the real Hilbert space $L_{\tilde{M}}$.

\subsection{State spaces}

To each hypersurface $\Sigma$ we associate a Hilbert space $\cH_\Sigma$ as follows. We consider the inner product $(\cdot,\cdot)\defeq\frac{1}{2}\{\cdot,\cdot\}_\Sigma$ on $L_\Sigma$. According to Section~\ref{sec:gint} this yields a $\sigma$-algebra $\cM_\Sigma^*$ and a Gaussian measure $\nu_\Sigma$ on $\hat{L}_\Sigma$. We def\/ine $\cH_\Sigma\defeq\rH^2(\hat{L}_\Sigma,\nu_\Sigma)$ as described in Section~\ref{sec:hcoh} as the space of square-integrable holomorphic functions on $\hat{L}_\Sigma$ with inner product (\ref{eq:hip}). When convenient, we also make use of the fact that, according to Theorem~\ref{thm:h2holom}, we may view $\cH_\Sigma$ as a~subspace of $\rH(L_\Sigma)$. In particular, this implies that an element of $\cH_\Sigma$ is completely determined by its values on $L_\Sigma$.

Note that a function that is holomorphic on $L_\Sigma$ is anti-holomorphic on $L_{\overline{\Sigma}}$ and vice-versa. In particular, complex conjugation def\/ines a conjugate linear isomorphism $\iota_\Sigma:\cH_\Sigma\to\cH_{\overline{\Sigma}}$. It is also clear that for disjoint unions of hypersurfaces we get $\cH_{\Sigma_1\cup\Sigma_2}=\cH_{\Sigma_1}\ctens\cH_{\Sigma_2}$, where the tensor product is the (completed) tensor product of Hilbert spaces. Thus, we have satisf\/ied axioms (T1), (T1b), (T2), (T2b).

Following Section~\ref{sec:hcoh} we consider coherent states in $\cH_\Sigma$. The coherent state corresponding to an element $\xi\in L_\Sigma$ is the holomorphic function
\begin{gather*}
 K_\xi(\phi)\defeq\exp\left(\frac{1}{2}\{\xi, \phi\}_\Sigma\right)\qquad\forall\, \phi\in\hat{L}_\Sigma .
\end{gather*}
Here, analogous to the corresponding context of Section~\ref{sec:hcoh}, we denote by $\{\cdot,\cdot\}_\Sigma$ not only the inner product on $L_\Sigma$, but also its extensions to sesquilinear maps $L_\Sigma\times\hat{L}_\Sigma$ and $\hat{L}_\Sigma\times L_\Sigma$.
Recall key properties of coherent states, due to Propositions~\ref{prop:csinf} and \ref{prop:coheval},
\begin{gather}
\langle K_\xi,\psi\rangle_\Sigma=\psi(\xi)\qquad\forall\,\xi\in L_\Sigma,\quad\forall\, \psi\in\cH_\Sigma,\nonumber\\
\langle K_\xi,K_{\xi'}\rangle_\Sigma=\exp\left(\frac{1}{2}\{\xi',\xi\}_\Sigma\right)\qquad
 \forall\, \xi,\xi'\in L_\Sigma,\nonumber\\
\|K_\xi\|_2= \exp\left(\frac{1}{4}\|\xi\|_{\Sigma,2}^2\right)\qquad\forall\, \xi\in L_\Sigma,\nonumber\\
 \langle\psi,\eta\rangle_\Sigma=\int_{\hat{L}_\Sigma} \langle\psi,K_\xi\rangle_\Sigma
  \langle K_\xi,\eta\rangle_\Sigma\,
 \xd\nu_\Sigma(\xi) \qquad\forall\, \psi,\eta\in\cH_\Sigma .
\label{eq:cohcompl}
\end{gather}
An important variant of the coherent states are their normalized versions, for $\xi\in L_\Sigma$
\begin{gather}
 \tilde{K}_\xi\defeq C_\xi K_\xi,\qquad\text{with}\quad C_\xi\defeq\exp\left(-\frac{1}{4}\{\xi, \xi\}_\Sigma\right) .
\label{eq:ncoh}
\end{gather}
When it is useful, we also indicate explicitly on which hypersurface a coherent state lives, e.g., we write $K_{\Sigma,\xi}$ to indicated this. Coherent states are compatible with the involutions $\iota_{\Sigma}:\cH_{\Sigma}\to\cH_{\overline{\Sigma}}$ in the obvious way
\begin{gather*}
K_{\overline{\Sigma},\xi}=\iota_{\Sigma}(K_{\Sigma,\xi}) ,\qquad \tilde{K}_{\overline{\Sigma},\xi}=\iota_{\Sigma}(\tilde{K}_{\Sigma,\xi}) .
\end{gather*}
The coherent states are also compatible with decompositions of hypersurfaces in a simple way. Namely, for $(\xi,\xi')\in L_{\Sigma_1}\times L_{\Sigma_2}$ we have
\begin{gather*}
 K_{(\xi,\xi')}=K_\xi\tens K_{\xi'} ,
\end{gather*}
and analogously for the normalized coherent states
\begin{gather}
 \tilde{K}_{(\xi,\xi')}=\tilde{K}_\xi\tens \tilde{K}_{\xi'} \qquad\text{and}\qquad
 C_{(\xi,\xi')}=C_\xi C_{\xi'} .
\label{eq:decncoh}
\end{gather}

\subsection{Amplitudes}

Let $M$ be a region. Consider the Gaussian probability measure $\nu_{\tilde{M}}$ on $\hat{L}_{\tilde{M}}\subseteq \hat{L}_{\partial M}$ determined by the real inner product $(\cdot,\cdot)=\frac{1}{4}g_{\partial M}(\cdot,\cdot)$ restricted to $L_{\tilde{M}}$, according to Section~\ref{sec:gint}.
Let $\psi\in\cH_{\partial M}$ and def\/ine $\tilde{\psi}:\hat{L}_{\tilde{M}}\to\C$ by $\tilde{\psi}\defeq \psi\circ \hat{r}_{\tilde{M}}$, where we think of $\psi$ as an element of $\cL^2(\hat{L}_{\partial M},\nu_{\partial M})$ and $\hat{r}_{\tilde{M}}$ is the continuous extension of $r_{\tilde{M}}$. If $\tilde{\psi}\in\cL^1(\hat{L}_{\tilde{M}},\nu_{\tilde{M}})$ we def\/ine
\begin{gather}
 \rho_M(\psi)\defeq\int_{\hat{L}_{\tilde{M}}} \tilde{\psi}(\phi)\,\xd\nu_{\tilde{M}}(\phi) .
\label{eq:aint}
\end{gather}
As we shall see in a moment there is at least a dense subspace of elements of $\cH_{\partial M}$ which yield integrable functions in this sense, namely the linear combinations of coherent states. Thus, we satisfy axiom (T4). It is also clear that formula (\ref{eq:aint}) automatically satisf\/ies axiom (T5a) since the measure for a disjoint union of regions is the product measure. To simplify notation we shall in the following not write explicitly a map $\tau$, when composed with $\rho$.

Remarkably, knowing the decomposition of the boundary solution space it is possible to give an \emph{explicit} formula for the evaluation of a coherent state with the amplitude map.
\begin{prop}
\label{prop:cohampl}
Let $\xi\in L_{\partial M}$. Let $\xi=\xi^{\text{R}}+J_{\partial M} \xi^{\text{I}}$ be the decomposition of $\xi$ with respect to the orthogonal direct sum $L_{\partial M}=L_{\tilde{M}}\oplus J_{\partial M} L_{\tilde{M}}$ according to Lemma~{\rm \ref{lem:decis}}. Then,
\begin{gather}
 \rho_M(K_\xi)=\exp\left(\frac{1}{4} g_{\partial M}(\xi^{\text{R}},\xi^{\text{R}})-\frac{1}{4} g_{\partial M}(\xi^{\text{I}},\xi^{\text{I}})-\frac{\im}{2} g_{\partial M}(\xi^{\text{R}},\xi^{\text{I}})\right) .
\label{eq:cohampl}
\end{gather}
\end{prop}
\begin{proof}
For $\phi\in\hat{L}_{\tilde{M}}$ we can rewrite the coherent state function as follows
\begin{gather}
 K_\xi(\phi)=\exp\left(\frac{1}{2}\{\xi^{\text{R}}+J_{\partial M} \xi^{\text{I}}, \phi\}_{\partial M}\right)
=\exp\left(\frac{1}{2}\{\xi^{\text{R}}, \phi\}_\Sigma-\frac{\im}{2}\{\xi^{\text{I}}, \phi\}_{\partial M}\right)\nonumber\\
\hphantom{K_\xi(\phi)}{}
=\exp\left(\frac{1}{2}g_{\partial M}(\xi^{\text{R}}, \phi)-\frac{\im}{2}g_{\partial M}(\xi^{\text{I}}, \phi)\right) .
\label{eq:cohrest}
\end{gather}
Here we have used the fact that $\omega_{\partial M}$ vanishes on $L_{\tilde{M}}$ since $L_{\tilde{M}}$ is an isotropic subspace of $L_{\partial M}$. We see that $K_\xi$ is an almost translation invariant function on $L_{\tilde{M}}$. Indeed, it is translation invariant with respect to the subspace of $L_{\partial M}$ given by the orthogonal complement of the at most two-dimensional space spanned by $\xi^{\text{R}}$ and $\xi^{\text{I}}$. We can thus apply Proposition~\ref{prop:evalgint} to evaluate the integral (\ref{eq:aint}) as a simple two-dimensional Gaussian integral. This yields the stated result.

Nevertheless, let us remark that the calculation can be simplif\/ied even more. Suppose we replace the factor $\im$ in the second summand of the exponent in (\ref{eq:cohrest}) by a complex parame\-ter~$z$. Then, clearly, the expression is holomorphic in $z$. Moreover, the integral (\ref{eq:aint}) will also be holomorphic in~$z$. This can be easily seen by recalling that a function is holomorphic in $z$ near a point $p$ if and only if any complex integral along a closed loop in a small neighborhood of $p$ vanishes. Now, Fubini's theorem applies here and we can interchange the integrals. Now recall that an entire holomorphic (as in this case) is completely determined by its values on $\R$. It is therefore suf\/f\/icient to perform the integral~(\ref{eq:aint}) merely for $z\in\R$. But then, the integrand is almost translation invariant with respect to the subspace of $L_{\partial M}$ given by the orthogonal complement of the at most one-dimensional space spanned by $\xi^{\text{R}}-z\xi^{\text{I}}$. Its calculation requires the evaluation merely of a one-dimensional Gaussian integral. Indeed, the result is
\begin{gather*}
\exp\left(\frac{1}{4}g_{\partial M}(\xi^{\text{R}}-z \xi^{\text{I}}, \xi^{\text{R}}-z \xi^{\text{I}})\right),
\end{gather*}
leading to (\ref{eq:cohampl}).
\end{proof}

This result has a simple, but compelling physical interpretation. To see this more clearly, we use normalized states. For a normalized coherent state equation (\ref{eq:cohampl}) becomes
\begin{gather}
 \rho_M(\tilde{K}_\xi)=\exp\left(-\frac{1}{2} g_{\partial M}\big(\xi^{\text{I}},\xi^{\text{I}}\big)-\frac{\im}{2} g_{\partial M}\big(\xi^{\text{R}},\xi^{\text{I}}\big)\right) .
\label{eq:ncohampl}
\end{gather}
If we think in classical terms, the component $\xi^{\text{R}}$ of the boundary solution $\xi$ can be continued consistently to the interior and is hence classically allowed. The component $\xi^{\text{I}}$ does not posses such a continuation and is hence classically forbidden. This is ref\/lected precisely in equation~(\ref{eq:ncohampl}). If the classically forbidden component is not present, the amplitude is simply equal to~$1$. On the other hand, the presence of a classically forbidden component leads to an exponential suppression, governed precisely by the ``size'' of this component (measured in terms of the metric~$g_{\partial M}$). There appears also a phase depending on the ``mixing'' of the components, if both are present.

We now turn to the context of axiom (T3x). Let $\Sigma$ be a hypersurface. Then $\Sigma$ def\/ines an \emph{empty region} $\hat{\Sigma}$ with boundary $\partial\hat{\Sigma}=\overline{\Sigma}\cup\Sigma'$. Here, $\Sigma'$ denotes a second copy of $\Sigma$. We have then, $L_{\partial \hat{\Sigma}}=L_{\overline{\Sigma}}\times L_{\Sigma'}$ and  $\cH_{\partial \hat{\Sigma}}=\cH_{\overline{\Sigma}}\ctens\cH_{\Sigma'}$. Moreover,
\begin{gather*}
 \omega_{\partial \hat{\Sigma}}=\omega_{\overline{\Sigma}}+\omega_{\Sigma'}
  =\omega_{\Sigma'}-\omega_{\Sigma},\qquad
 J_{\partial \hat{\Sigma}}=J_{\overline{\Sigma}}+J_{\Sigma'}
 =J_{\Sigma'}-J_{\Sigma},\\
 g_{\partial \hat{\Sigma}}=g_{\overline{\Sigma}}+g_{\Sigma'}=g_{\Sigma}+g_{\Sigma'},\\
 \{\cdot,\cdot\}_{\partial \hat{\Sigma}}
 =\{\cdot,\cdot\}_{\overline{\Sigma}}+\{\cdot,\cdot\}_{\Sigma'}
 =\overline{\{\cdot,\cdot\}_{\Sigma}}
  +\{\cdot,\cdot\}_{\Sigma'} .
\end{gather*}
The subspace $L_{\tilde{\hat{\Sigma}}}\subseteq L_{\partial \hat{\Sigma}}$ is precisely the space of pairs $(\phi,\phi)$ while the subspace $J_{\partial \hat{\Sigma}} L_{\tilde{\hat{\Sigma}}}\subseteq L_{\partial \hat{\Sigma}}$ is the space of pairs $(\phi,-\phi)$. In particular, we have a linear bijection of real vector spaces $L_\Sigma\to L_{\tilde{\hat{\Sigma}}}$ given by $\phi\mapsto (\phi,\phi)$. The following proposition shows that axiom (T3x) is satisf\/ied.

\begin{prop}
\label{prop:ipampl}
We have $\cH_{\overline{\Sigma}}\tens\cH_{\Sigma'}\subseteq\cH_{\partial \hat{\Sigma}}^\circ$. Moreover, for $\psi,\psi'\in \cH_\Sigma$ we have
\begin{gather}
 \rho_{\hat{\Sigma}}(\iota_\Sigma(\psi)\tens\psi')
=\langle\psi, \psi'\rangle_\Sigma .
\label{eq:ipampl}
\end{gather}
\end{prop}
\begin{proof}
Under the bijection $\phi\mapsto (\phi,\phi)$ the quadratic forms determined by  $\frac{1}{2}\{\cdot,\cdot\}_\Sigma$ on $L_\Sigma$ and by $\frac{1}{4}g_{\partial \hat{\Sigma}}$ on $L_{\tilde{\hat{\Sigma}}}\subseteq L_{\partial \hat{\Sigma}}$ become identical
\begin{gather*}
 \frac{1}{4} g_{\partial \hat{\Sigma}}(\phi,\phi)=\frac{1}{4} g_\Sigma(\phi,\phi)
 +\frac{1}{4} g_{\Sigma'}(\phi,\phi)=\frac{1}{2} g_\Sigma(\phi,\phi)
 =\frac{1}{2} \{\phi,\phi\}_\Sigma .
\end{gather*}
This implies the equality of the measures $\nu_\Sigma$ and $\nu_{\tilde{\hat{\Sigma}}}$ under this identif\/ication. This in turn implies the equality for all $\psi,\psi'\in\cH_\Sigma$
\begin{gather}
\int_{\hat{L}_{\tilde{\hat{\Sigma}}}} \overline{\psi(\phi)}\psi'(\phi)\,\xd\nu_{\tilde{\hat{\Sigma}}}(\phi)
=\int_{\hat{L}_\Sigma} \overline{\psi(\phi)}\psi'(\phi)\,\xd\nu_{\Sigma}(\phi)
=\langle\psi, \psi'\rangle_\Sigma .
\label{eq:eqmes}
\end{gather}
Finally notice that by def\/inition of the amplitude map
\begin{gather*}
 \rho_{\hat{\Sigma}}(\iota_\Sigma(\psi)\tens\psi')
=\int_{\hat{L}_{\tilde{\hat{\Sigma}}}} \overline{\psi(\phi)}\psi'(\phi)\,\xd\nu_{\tilde{\hat{\Sigma}}}(\phi)
\end{gather*}
if the integrand is integrable. But this is precisely ensured by (\ref{eq:eqmes}). Hence, $\cH_{\overline{\Sigma}}\tens\cH_{\Sigma'}\subseteq\cH_{\partial \hat{\Sigma}}^\circ$ and equation (\ref{eq:ipampl}) is valid for all $\psi,\psi'\in\cH_\Sigma$.
\end{proof}

\subsection{Gluing}

We proceed in this section to demonstrate the validity of the gluing axiom (T5b). In order to do this we will need to introduce an additional assumption. First, however, we shall modify the appearance of the axiom to a more convenient form. For a hypersurface $\Sigma$ we can write the completeness relation in the Hilbert space $\cH_\Sigma$ either with an orthonormal basis, or with coherent states as in Proposition~\ref{prop:coheval}.
Implicit in axiom (T5b) as written in Section~\ref{sec:coreaxioms} is the f\/irst form with an orthonormal basis. We shall prefer here the second form and shall require instead of equation (\ref{eq:glueax1}), for all $\psi\in\cH_{\Sigma_1}^\ds$
\begin{gather}
 \rho_{M_1}(\psi)\cdot c(M;\Sigma,\overline{\Sigma'})=\int_{\hat{L}_\Sigma}\rho_M(\psi\tens K_\xi\tens\iota_\Sigma(K_\xi))\,\xd\nu_\Sigma(\xi) .
\label{eq:glueax2}
\end{gather}
Recall that here $M$ is a region with its boundary decomposing as a disjoint union $\partial M=\Sigma_1\cup\Sigma\cup \overline{\Sigma'}$, where $\Sigma'$ is a copy of $\Sigma$. $M_1$ denotes the gluing of $M$ with itself along $\Sigma$, $\overline{\Sigma'}$ and we suppose that $M_1$ is an admissible region. We note $\partial M_1=\Sigma_1$. Then, we need (\ref{eq:glueax2}) to be satisf\/ied. However, taking merely the geometric setting of Section~\ref{sec:geomax} and the classical axioms of Section~\ref{sec:classax} is not suf\/f\/icient to ensure this\footnote{One can construct counter-examples along the lines of simple topological quantum f\/ield theories where gluing a cylinder to itself yields a torus and associated to it the dimension of the vector space associated with one boundary component. If the dimension of the vector space is inf\/inite, this becomes ill def\/ined.}.

\begin{dfn}
We say that the gluing data satisfy the \emph{integrability condition} if
the function
\begin{gather*}
 \xi\mapsto \rho_M(K_0\tens K_\xi\tens\iota_\Sigma(K_\xi))
\end{gather*}
is integrable with respect to $(\hat{L}_\Sigma,\nu_{\Sigma})$ and its integral is dif\/ferent from zero.
\end{dfn}

We shall introduce the additional assumption that the integrability condition is satisf\/ied for any gluing yielding an admissible region. Then, all axioms are indeed satisf\/ied as the following theorem shows.

\begin{thm}
If the integrability condition is satisfied, then axiom {\rm (T5b)} holds. Moreover,
\begin{gather*}
c(M;\Sigma,\overline{\Sigma'})
=\int_{\hat{L}_\Sigma}\rho_M(K_0\tens K_\xi\tens\iota_\Sigma(K_\xi))\,
 \xd\nu_\Sigma(\xi) .
\end{gather*}
\end{thm}
\begin{proof}
Since the space of linear combinations of coherent states is dense in $\cH_{\partial M_1}^\circ$ it is suf\/f\/icient to demonstrate the axiom for coherent states. Thus, we have to show for all $\phi\in L_{\Sigma_1}$
\begin{gather}
 \rho_{M_1}(K_{\phi})\cdot c(M;\Sigma,\overline{\Sigma'})=\int_{\hat{L}_\Sigma}\rho_M(K_{\phi}\tens K_\xi\tens\iota_\Sigma(K_\xi))\,\xd\nu_\Sigma(\xi) .
\label{eq:gluecoh}
\end{gather}
We f\/irst restrict the proof to the special case $\phi\in L_{\tilde{M}_1}$. Then, by (C7) there is $\tilde{\phi}\in L_{\tilde{M}}$ such that $\tilde{\phi}_1=\phi$ (with the obvious notation). Moreover, also due to (C7) we have $\tilde{\phi}_\Sigma=\tilde{\phi}_{\Sigma'}$.

We start by reformulating the integrand of (\ref{eq:gluecoh}). We perform a substitution using Proposition~\ref{prop:rtransinv} with $x=\tilde{\phi}$, use the fact that $L_{\tilde{M}}$ is Lagrangian in $L_{\partial{M}}$, and make use of various identities relating inner products, symplectic forms etc.\ that follow straightforwardly from the axioms for the classical data to obtain
\begin{gather*}
  \rho_M(K_{\phi}\tens K_\xi\tens\iota_\Sigma(K_\xi))
   = \int_{\hat{L}_{\tilde{M}}} K_{\phi}(\eta_1)K_\xi(\eta_\Sigma)
 \overline{K_{\xi}(\eta_{\Sigma'})}\,\xd\nu_{\tilde{M}}(\eta)\\
 \qquad{}  = \int_{\hat{L}_{\tilde{M}}} \exp\left(\frac{1}{2}\{ \phi,\eta_1\}_1
 +\frac{1}{2}\{ \xi,\eta_\Sigma\}_\Sigma
 +\frac{1}{2}\{ \xi,\eta_{\Sigma'}\}_{\overline{\Sigma'}}\right)
 \xd\nu_{\tilde{M}}(\eta)\\
\qquad{}   = \int_{\hat{L}_{\tilde{M}}} \exp\left(\frac{1}{2}\{ \phi,\eta_1+\tilde{\phi}_1\}_1
 +\frac{1}{2}\{ \xi,\eta_\Sigma+\tilde{\phi}_{\Sigma}\}_\Sigma
 +\frac{1}{2}\{ \xi,\eta_{\Sigma'}+\tilde{\phi}_{\Sigma'}\}_{\overline{\Sigma'}}\right.\\
 \left.\quad\qquad{} -\frac{1}{4}g_{\partial M}(\tilde{\phi}+2\eta,\tilde{\phi})
  \right)\xd\nu_{\tilde{M}}(\eta)\\
 \qquad{}  = \int_{\hat{L}_{\tilde{M}}} \exp\left(\frac{1}{2}\{ \phi,\eta_1+\tilde{\phi}_1\}_1
 +\frac{1}{2}\{ \xi,\eta_\Sigma+\tilde{\phi}_{\Sigma}\}_\Sigma
 +\frac{1}{2}\{ \xi,\eta_{\Sigma'}+\tilde{\phi}_{\Sigma'}\}_{\overline{\Sigma'}}\right.\\
 \left.\quad\qquad{}  -\frac{1}{4}\{\tilde{\phi},\tilde{\phi}+2\eta\}_{\partial M}
  \right)\xd\nu_{\tilde{M}}(\eta)\\
\qquad{}  = \int_{\hat{L}_{\tilde{M}}} \exp\left(\frac{1}{4}\{ \phi,\phi\}_1
 +\frac{1}{2}\{ \xi-\tilde{\phi}_\Sigma,\eta_\Sigma\}_\Sigma
 +\frac{1}{2}\{ \xi-\tilde{\phi}_{\Sigma'},\eta_{\Sigma'}\}_{\overline{\Sigma'}}\right.\\
\left.
 \quad\qquad{} +\frac{1}{4}\{ 2\xi-\tilde{\phi}_\Sigma,\tilde{\phi}_\Sigma\}_\Sigma
 +\frac{1}{4}\{ 2\xi-\tilde{\phi}_{\Sigma'},\tilde{\phi}_{\Sigma'}\}_{\overline{\Sigma'}}
 \right)\xd\nu_{\tilde{M}}(\eta)\\
 \qquad{} = \exp\left(\frac{1}{4}\{ \phi,\phi\}_1
 +\frac{1}{4}\{ 2\xi-\tilde{\phi}_\Sigma,\tilde{\phi}_\Sigma\}_\Sigma
 +\frac{1}{4}\{ 2\xi-\tilde{\phi}_{\Sigma'},\tilde{\phi}_{\Sigma'}\}_{\overline{\Sigma'}}\right)\\
 \quad\qquad{}\times \int_{\hat{L}_{\tilde{M}}} \exp\left(
 \frac{1}{2}\{ \xi-\tilde{\phi}_\Sigma,\eta_\Sigma\}_\Sigma
 +\frac{1}{2}\{ \xi-\tilde{\phi}_{\Sigma'},\eta_{\Sigma'}\}_{\overline{\Sigma'}}\right)\xd\nu_{\tilde{M}}(\eta)\\
 \qquad{} = \exp\left(\frac{1}{4}\{ \phi,\phi\}_1
 +\frac{1}{2} g_\Sigma(2\xi-\tilde{\phi}_\Sigma,\tilde{\phi}_\Sigma)
 \right)\\
 \quad\qquad{}\times \int_{\hat{L}_{\tilde{M}}} \exp\left(
 \frac{1}{2}g_\Sigma(\xi-\tilde{\phi}_\Sigma,\eta_\Sigma+\eta_{\Sigma'})
 +\im\omega_\Sigma(\xi-\tilde{\phi}_\Sigma,\eta_\Sigma-\eta_{\Sigma'})\right)\xd\nu_{\tilde{M}}(\eta) .
\end{gather*}
We now integrate the expression over $\xi\in\hat{L}_\Sigma$ and use Proposition~\ref{prop:rtransinv} with $x=-\tilde{\phi}$ to obtain
\begin{gather*}
  \int_{\hat{L}_\Sigma} \left(\exp\left(\frac{1}{4}\{ \phi,\phi\}_1
 -\frac{1}{2} g_\Sigma(2\xi-\tilde{\phi}_\Sigma,-\tilde{\phi}_\Sigma)
 \right)\right. \\
\left. \qquad{}\times \int_{\hat{L}_{\tilde{M}}} \exp\left(
 \frac{1}{2}g_\Sigma(\xi-\tilde{\phi}_\Sigma,\eta_\Sigma+\eta_{\Sigma'})
 +\im\omega_\Sigma(\xi-\tilde{\phi}_\Sigma,\eta_\Sigma-\eta_{\Sigma'})\right)\xd\nu_{\tilde{M}}(\eta)\right)\xd\nu_\Sigma(\xi)\\
  = \int_{\hat{L}_\Sigma} \left(\exp\left(\frac{1}{4}\{ \phi,\phi\}_1
  \right)
\int_{\hat{L}_{\tilde{M}}} \exp\left(
 \frac{1}{2}g_\Sigma(\xi,\eta_\Sigma+\eta_{\Sigma'})
 +\im\omega_\Sigma(\xi,\eta_\Sigma-\eta_{\Sigma'})\right)\xd\nu_{\tilde{M}}(\eta)\right)\xd\nu_\Sigma(\xi)\\
  = \exp\left(\frac{1}{4}\{ \phi,\phi\}_1\right)
 \int_{\hat{L}_\Sigma}\rho_M(K_0\tens K_\xi\tens\iota_\Sigma(K_\xi))\,
 \xd\nu_\Sigma(\xi)  = \rho_{M_1}(K_\phi) \cdot c(M;\Sigma,\overline{\Sigma'}) .
\end{gather*}

For the case of general $\phi\in L_{\Sigma_1}$ consider the canonical decomposition $\phi=\phi^{\text{R}}+J_{\Sigma_1} \phi^{\text{I}}$, where $\phi^{\text{R}},\phi^{\text{I}}\in L_{\tilde{M_1}}$. Write
\begin{gather*}
K_\phi(\eta_1)=\exp\left(\frac{1}{2}\{ \phi^{\text{R}}+J_{\Sigma_1}\phi^{\text{I}},\eta_1\}_1\right)=
 \exp\left(\frac{1}{2}\{\phi^{\text{R}},\eta_1\}_1-\frac{\im}{2}\{\phi^{\text{I}},\eta_1\}_1\right)
\end{gather*}
Then, replace $\im$ in the second summand of the exponential by a complex parameter $z$. Observe that we obtain a holomorphic function of $z$ and that the integrations to be performed preserve the holomorphicity as in the proof of Proposition~\ref{prop:cohampl}. Hence, it is suf\/f\/icient to perform the calculation for $z\in\R$ to determine the function completely. But this is precisely the calculation performed above with $\phi$ replaced by $\phi^{\text{R}}-z \phi^{\text{I}}\in L_{\tilde{M_1}}$. This completes the proof.
\end{proof}

\subsection{Evolution picture}
\label{sec:evol}

In this section we consider what our quantization scheme implies in an ``evolution picture''. That is, we consider situations with regions were there is a one-to-one correspondence between classical solutions on one boundary component and those on another boundary component. This is useful in particular for comparison with other quantization schemes where (time-)evolution plays a distinguished role.

Let $M$ be a region such that its boundary decomposes as a disjoint union of two components $\partial M=\Sigma_1\cup\overline{\Sigma_2}$. Assume moreover, that the linear maps $r_1:L_{\tilde{M}}\to L_{\Sigma_1}$ and $r_2:L_{\tilde{M}}\to L_{\Sigma_2}$ given by $r_{M;\Sigma_1,\overline{\Sigma_2}}$ composed with orthogonal projections are homeomorphisms. We denote the composition by $T\defeq r_2\circ r_1^{-1}:L_{\Sigma_1}\to L_{\Sigma_2}$. As explained in Section~\ref{sec:classft} we have
\begin{gather}
\omega_{\Sigma_2}(T\phi,T\phi')=\omega_{\Sigma_1}(\phi,\phi')\qquad\forall\,\phi,\phi'\in L_{\Sigma_1}
\label{eq:symevol}
\end{gather}
due to axiom (C5). However, we do not necessarily have
\begin{gather}
J_{\Sigma_2}\circ T=T\circ J_{\Sigma_1} .
\label{eq:jevol}
\end{gather}
But if (and only if) this is true, $T$ is unitary and we obtain a particularly ``nice'' evolution picture.

\begin{prop}
\label{prop:evol}
There is a linear map $U:\cH_{\Sigma_1}\to\cH_{\Sigma_2}$ such that
\begin{gather}
 \rho_M(\psi_1\tens\iota_{\Sigma_2}(\psi_2))=\langle \psi_2,U\psi_1\rangle_{\Sigma_2}\qquad
 \forall \, \psi_1,\psi_2\in\cH_{\Sigma_2}.
\label{eq:ipamplevol}
\end{gather}
In particular, $U$ is given by
\begin{gather}
(U\psi)(\phi)=\rho_M\big(\psi\tens K_{\overline{\Sigma_2},\phi}\big)
 \qquad\forall \, \psi\in\cH_{\Sigma_1},\quad
  \forall\,\phi\in L_{\Sigma_2}.
\label{eq:stateevol}
\end{gather}
Moreover, if $T$ is unitary then $U$ is unitary and we have
\begin{gather*}
(U\psi)(\phi)=\psi\big(T^{-1}\phi\big)\qquad\forall \,\psi\in\cH_{\Sigma_1}, \quad
  \forall\,\phi\in L_{\Sigma_2},\\
U K_{\Sigma_1,\xi}=K_{\Sigma_2,T\xi}\qquad \forall\,\xi\in L_{\Sigma_1} .
\end{gather*}
\end{prop}

\begin{proof}
Suppose we take (\ref{eq:stateevol}) as a def\/inition. It is then straightforward to verify (\ref{eq:ipamplevol}) using the formula (\ref{eq:aint}) for the amplitude, the completeness relation (\ref{eq:cohcompl}) and further properties of coherent states:
\begin{gather*}
  \langle\psi_2,U\psi_1\rangle_{\Sigma_2}
  = \int_{\hat{L}_{\Sigma_2}} \overline{\psi_2(\phi)}\,
  \rho_M\big(\psi_1\tens K_{\overline{\Sigma_2},\phi}\big)
 \xd\nu_{\Sigma_2}(\phi) \\
\hphantom{\langle\psi_2,U\psi_1\rangle_{\Sigma_2}}{}
  = \int_{\hat{L}_{\Sigma_2}} \overline{\psi_2(\phi)}\left(\int_{\hat{L}_{\tilde{M}}}
 \psi_1(\eta) K_{\overline{\Sigma_2},\phi}(\eta)\,
 \xd\nu_{\tilde{M}}(\eta)\right)\xd\nu_{\Sigma_2}(\phi) \\
 \hphantom{\langle\psi_2,U\psi_1\rangle_{\Sigma_2}}{}
  = \int_{\hat{L}_{\tilde{M}}} \psi_1(\eta)\left(\int_{\hat{L}_{\Sigma_2}}
 \overline{\psi_2(\phi)} K_{\Sigma_2,\eta}(\phi)\,\xd\nu_{\Sigma_2}(\phi)
 \right)\xd\nu_{\tilde{M}}(\eta) \\
 \hphantom{\langle\psi_2,U\psi_1\rangle_{\Sigma_2}}{}
  = \int_{\hat{L}_{\tilde{M}}} \psi_1(\eta)
 \overline{\psi_2(\eta)}\,\xd\nu_{\tilde{M}}(\eta)
  = \rho_M(\psi_1\tens\iota_{\Sigma_2}(\psi_2))
\end{gather*}

We proceed to consider the special case that $T$ is unitary. Then, we have an identif\/ication of the real vector spaces $L_{\tilde{M}}$ and $L_{\Sigma_1}$ such that $\nu_{\tilde{M}}=\nu_{\Sigma_1}$ as in the proof of Proposition~\ref{prop:ipampl}. Explicitly, we have
\begin{gather*}
  \rho_M\big(\psi\tens K_{\overline{\Sigma_2},\phi}\big)
  = \int_{\hat{L}_{\tilde{M}}} \psi(\eta_1)
 K_{\overline{\Sigma_2},\phi}(\eta_2)\,\xd\nu_{\tilde{M}}(\eta)
  = \int_{\hat{L}_{\tilde{M}}} \psi(\eta_1)
 \exp\left(\frac{1}{2}\{\phi,\eta_2\}_{\overline{\Sigma_2}}\right)
\xd\nu_{\tilde{M}}(\eta) \\
\hphantom{\rho_M\big(\psi\tens K_{\overline{\Sigma_2},\phi}\big)}{}
  = \int_{\hat{L}_{\tilde{M}}} \psi(\eta_1)
 \overline{\exp\left(\frac{1}{2}\big\{T^{-1}\phi,T^{-1}\eta_2\big\}_{\Sigma_1}\right)}
\xd\nu_{\tilde{M}}(\eta) \\
\hphantom{\rho_M\big(\psi\tens K_{\overline{\Sigma_2},\phi}\big)}{}
  = \int_{\hat{L}_{\tilde{M}}} \psi(\eta_1)
 \overline{K_{\Sigma_1,T^{-1}\phi}(\eta_1)}\,\xd\nu_{\tilde{M}}(\eta) \\
 \hphantom{\rho_M\big(\psi\tens K_{\overline{\Sigma_2},\phi}\big)}{}
  = \int_{\hat{L}_{\Sigma_1}} \psi(\eta)
 \overline{K_{\Sigma_1,T^{-1}\phi}(\eta)}\,\xd\nu_{\Sigma_1}(\eta)
  = \langle K_{\Sigma_1,T^{-1}\phi},\psi\rangle_{\Sigma_1}
  = \psi\big(T^{-1}\phi\big) .
\end{gather*}
Here, we have used the notation $\eta_1$ and $\eta_2$ to specify explicitly that we refer to the solutions induced on the hypersurfaces $\Sigma_1$ and $\Sigma_2$ respectively. The explicit expression for the evolved coherent state is an easy calculation.
\end{proof}

Consider the following particular physical application of this statement. Suppose we are in a setting where the spaces of solutions $L_{\Sigma_1}$, $L_{\Sigma_2}$ for any pair $(\Sigma_1,\Sigma_2)$ of connected admissible hypersurfaces can be related by linear homeomorphisms $T_{\Sigma_1,\Sigma_2}:L_{\Sigma_1}\to L_{\Sigma_2}$ as described above, either directly or indirectly. By directly, we mean that there is a region with boundary the disjoint union $\Sigma_1\cup\overline{\Sigma_2}$. By indirect we mean that there is a third hypersurface $\Sigma_3$, such that both $L_{\Sigma_1}$ and $L_{\Sigma_2}$ can be directly related to $L_{\Sigma_3}$. Note that uniqueness of the maps $T_{\Sigma_1,\Sigma_2}$ is ensured by the axioms (C1)--(C7). The most important example for such a setting would be that of Cauchy hypersurfaces in a globally hyperbolic spacetime. As observed above, we have compatibility of the symplectic forms as written in equation (\ref{eq:symevol}). To def\/ine a complete quantization it now suf\/f\/ices to pick a compatible complex structure $J_{\Sigma_0}$ on one particular admissible hypersurface $\Sigma_0$. This can then be transported with the maps $T$ to any other admissible hypersurface via equation~(\ref{eq:jevol}). Clearly, by Proposition~\ref{prop:evol} we obtain a quantization where evolution $U_{\Sigma_1,\Sigma_2}:\cH_{\Sigma_1}\to\cH_{\Sigma_2}$ from any admissible hypersurface $\Sigma_1$ to any other admissible hypersurface $\Sigma_2$ is unitary. Note that this is in agreement with recent results in \cite{CoOe:unitary}, where a~Schr\"odinger--Feynman quantization was used instead.

\subsection{Vacuum}

The quantization construction described above yields a natural vacuum in the sense of the axioms of Section~\ref{sec:vacax}. Indeed, it is immediate to verify that def\/ining the vacuum state $\psi_{\Sigma,0}$ for every hypersurface $\Sigma$ to be the constant function with value $1$ satisf\/ies all the axioms. Alternatively, we can view this as the coherent state $K_0$ associated to the element $0\in L_\Sigma$.

There is a potentially larger class of vacua that also quite naturally arises from our quantization construction. To this end suppose that all regions and hypersurfaces arise as submanifolds of a f\/ixed manifold $B$ (possibly with additional structure) of dimension $d$. (This setting was termed a \emph{global background} in \cite{Oe:GBQFT}.) Suppose now that there exists a solution $\phi$ of the classical f\/ield equations in all of $B$. This induces a particular local solution in any region and on any hypersurface. The normalized coherent states associated with these solutions then form a vacuum in the sense of the axioms. We may formalize this as follows.

\begin{dfn}
\label{dfn:gsol}
Let $\{\phi_\Sigma\}$ be an assignment of an element $\phi_\Sigma\in L_\Sigma$ to every hypersurface $\Sigma$. Then we call this assignment a \emph{global solution} if\/f it satisf\/ies the following properties:
\begin{itemize}\itemsep=0pt
\item[1.] Let $\Sigma$ be a hypersurface. Then, $\phi_{\overline{\Sigma}}=\phi_{\Sigma}$.
\item[2.] Suppose the hypersurface $\Sigma$ decomposes into a disjoint union of hypersurfaces $\Sigma=\Sigma_1\cup\cdots\cup\Sigma_n$. Then, $\phi_\Sigma=(\phi_{\Sigma_1},\dots,\phi_{\Sigma_n})$.
\item[3.] Let $M$ be a region. Then, $\phi_{\partial M}\in L_{\tilde{M}}$.
\end{itemize}
\end{dfn}

Note that we have not explicitly f\/ixed any solutions in the spaces $L_M$ for regions $M$ as this is not necessary for our present purposes. Of course, the last condition in the def\/inition implies that the selected solution in the boundary of a region comes from a solution in the interior. Note also that we do not need to explicitly require a global background. In particular, we see that selecting the trivial solution on each hypersurface also satisf\/ies the def\/inition of a global solution.

\begin{prop}
Every global solution gives rise to a vacuum. In particular, if $\{\phi_\Sigma\}$ defines a global solution, the associated vacuum state $\psi_{\Sigma,0}$ for the hypersurface $\Sigma$ is the normalized coherent state $\tilde{K}_{\phi_\Sigma}$.
\end{prop}
\begin{proof}
(V1) is clear. Condition~1  of Def\/inition~\ref{dfn:gsol} together with axiom (C2) and the explicit form~(\ref{eq:ncoh}) of the normalized coherent state ensures (V2). Condition~2  of Def\/inition~\ref{dfn:gsol} together with axiom (C3) and the decomposition property~(\ref{eq:decncoh}) of coherent states yields~(V3). Finally combining condition~3  of Def\/inition~\ref{dfn:gsol}, Proposition~\ref{prop:cohampl} and (\ref{eq:ncoh}) yields~(V5).
\end{proof}

\section[Klein-Gordon theory in special geometries]{Klein--Gordon theory in special geometries}
\label{sec:kg}

In this section we consider the Klein--Gordon theory and its quantization in Minkowski space for certain special families of hypersurfaces and regions. This serves partly for comparison with previous results \cite{CoOe:spsmatrix,CoOe:smatrixgbf,Oe:KGtl, Oe:timelike}, where a quantization scheme based on the Schr\"odinger representation and the Feynman path integral was used. However, we are also able to extend these results in a certain direction. In particular, we show how classical solutions that have exponential behavior in spatial directions can be quantized on certain timelike hypersurfaces. These evanescent waves are ``invisible'' in many traditional quantization schemes as they do not yield well behaved global solutions.

The hypersurfaces we consider are essentially of three types: 1.~Equal-time hyperplanes, 2.~Timelike hyperplanes along the temporal axis, 3.~Timelike hypercylinders. The f\/irst geometry represents the most simple ``standard case'' of what one would usually consider in various well-known quantization schemes. A~quantization on the second geometry was f\/irst considered in~\cite{Oe:timelike} and on the third geometry in~\cite{Oe:KGtl}. We use coordinates $(t,x_1,x_2,x_3)$ on Minkowski space.

\subsection{Equal-time hyperplanes}

We consider the following geometrical setting: Admissible hypersurfaces are all oriented equal-time hyperplanes and their f\/inite disjoint unions. Admissible regular regions are all f\/inite closed time-intervals extended over all of space and their f\/inite disjoint unions.

We parametrize global solutions of the Klein--Gordon equation in the usual way via plane waves
\begin{gather}
 \phi(t,x)=\int\frac{\xd^3 k}{(2\pi)^3 2E}
  \left(\phi(k) e^{-\im(E t-k x)}+\overline{\phi(k)} e^{\im(E t-k x)}\right) ,
\label{eq:kgsolg}
\end{gather}
where $\phi:\R^3\to\C$ is a complex function on momentum space of a type to be determined below and $E\defeq\sqrt{k^2+m^2}$. Due to the usual Cauchy property we can use this also to parametrize the space of solutions $L_t$ in a neighborhood of an equal-time hypersurface $\Sigma_t$. The (negative of the) symplectic form (\ref{eq:sympl}) on $L_t$ is\footnote{\label{fn:nsym}It seems that to obtain results in agreement with \cite{CoOe:spsmatrix,CoOe:smatrixgbf,Oe:KGtl, Oe:timelike} one should chose \emph{minus} the symplectic form rather than the symplectic form itself. Changing simultaneously the sign of the complex structure, this leaves the axioms of Section~\ref{sec:classax} invariant and therefore may be considered merely a choice of convention for our present purposes.}
\begin{gather*}
 \omega_t(\phi_1,\phi_2)   =\frac{1}{2}\int\xd^3 x\,
  \left(\phi_2(t,x) \partial_0 \phi_1(t,x) - \phi_1(t,x)\partial_0\phi_2(t,x)\right) \nonumber\\
 \hphantom{\omega_t(\phi_1,\phi_2)}{}
  =\frac{\im}{2}\int\frac{\xd^3 k}{(2\pi)^3 2E}
 \left(\phi_2(k)\overline{\phi_1(k)}-\phi_1(k)\overline{\phi_2(k)}\right) .
\end{gather*}
In agreement with \cite{CoOe:spsmatrix,CoOe:smatrixgbf,Oe:KGtl, Oe:timelike} we chose the orientation of $\Sigma_t$ here as the past boundary of a region that lies in the future.
According to Section~\ref{sec:geomquant} we need to f\/ind a suitable complex structure on $L_t$. The standard one is
\begin{gather*}
 (J(\phi))(k)=-\im\phi(k) .
\end{gather*}
This complex structure is automatically compatible with time-evolution in the sense of (\ref{eq:jevol}), since we use a parametrization in terms of global solutions. Thus, we have unitary evolution in time according to Proposition~\ref{prop:evol}.

The induced metric (\ref{eq:kmetric}) is
\begin{gather*}
 g_t(\phi_1,\phi_2)= \int\frac{\xd^3 k}{(2\pi)^3 2E}
 \left(\phi_1(k)\overline{\phi_2(k)}+\phi_2(k)\overline{\phi_1(k)}\right) .
\end{gather*}
This leads to the standard complex inner product (\ref{eq:kcip}) on $L_t$
\begin{gather*}
 \{\phi_1,\phi_2\}_t = 2\int\frac{\xd^3 k}{(2\pi)^3 2E}\, \phi_1(k)\overline{\phi_2(k)} .
\end{gather*}
This also f\/ixes the space $L_t$ precisely as the space of (equivalence classes of) square-integrable complex functions $\phi:\R^3\to\C$ with the inner product written above. The quantization thus obtained with the unitary time evolution operator $U$ given by Proposition~\ref{prop:evol} is equivalent to the usual well-known quantization of Klein--Gordon theory.

The normalized coherent state $\tilde{K}_{t,\eta}$ takes in the Schr\"odinger representation the form
\begin{gather*}
 \tilde{K}_{t,\eta}(\varphi)
 =C_{\eta} \exp\left(\int\frac{\xd^3 x\, \xd^3 k}{(2\pi)^3}
  \eta(k) e^{-\im(E t-k x)}\varphi(x)\right)\psi_{0}(\varphi),
\end{gather*}
see equation (2) of \cite{CoOe:spsmatrix} and equation (26) of \cite{CoOe:smatrixgbf}.

We also note that the complex solution $\hat{\eta}$ of the Klein--Gordon equation appearing in equation~(6) of~\cite{CoOe:spsmatrix} and equation~(39) of \cite{CoOe:smatrixgbf} and determined by solutions $\eta_1\in L_{t_1}$ and $\eta_2\in L_{t_2}$ is simply $\hat{\eta}=\eta^{\text{R}}-\im\eta^{\text{I}}$ where we decompose $(\eta_1,\eta_2)\in L_{\partial[t_1,t_2]}$ (with the obvious notation) as in Lemma~\ref{lem:decis}, $(\eta_1,\eta_2)=\eta^{\text{R}}+ J_{\partial[t_1,t_2]} \eta^{\text{I}}$.

\subsection{Timelike hyperplanes}
\label{sec:kgtl}

We consider the following geometrical setting: Admissible hypersurfaces are all oriented hyperplanes perpendicular to the $x_1$-axis and their f\/inite disjoint unions. Admissible regular regions are all f\/inite closed intervals on the $x_1$-axis extended in the other coordinate directions and their f\/inite disjoint unions.

A solution of the Klein--Gordon equation that is well behaved near a constant $x_1$ hypersurface may be parametrized via a complex function $\eta:\R\times\R^2\to\C$ as
\begin{gather*}
 \eta(t,x_1,\tilde{x})=\int\frac{\xd E\, \xd^2 \tilde{k}}{(2\pi)^3 2k_1}
  \left(\eta(E,\tilde{k}) f(E,\tilde{k},x_1)
  e^{-\im(E t-\tilde{k} \tilde{x})}
+\overline{\eta(E,\tilde{k})}\, \overline{f(E,\tilde{k},x_1)}
  e^{\im(E t-\tilde{k} \tilde{x})}\right) ,
\end{gather*}
where $\tilde{x}\defeq(x_2,x_3)$, $\tilde{k}\defeq(k_2,k_3)$, $k_1\defeq \sqrt{|E^2-\tilde{k}^2-m^2|}$ and
\begin{gather*}
 f(E,\tilde{k},x_1)\defeq\begin{cases}
  e^{\im k_1 x_1}=\cos(k_1 x_1)+\im\sin(k_1 x_1) &
   \text{if}\; E^2-\tilde{k}^2-m^2>0, \\
   \cosh(k_1 x_1)+\im \sinh(k_1 x_1) & \text{if}\; E^2-\tilde{k}^2-m^2 <0 .
 \end{cases}
\end{gather*}
In contrast to the case of spacelike hypersurfaces the solutions now fall into two classes, depending on whether $E^2$ is larger or smaller than $\tilde{k}^2+m^2$. That is, they can be distinguished by whether the ``missing'' momentum-squared in the $x_1$-direction, $E^2-\tilde{k}^2-m^2$, is positive or negative. In the f\/irst case we obtain the usual plane waves that yield well behaved global solutions of the form~(\ref{eq:kgsolg}). But there are additional solutions corresponding to the second case that are well behaved near the hypersurface, but not globally. These are \emph{evanescent waves} that behave exponentially in the $x_1$-direction. In~\cite{Oe:KGtl, Oe:timelike} these were termed (somewhat unfortunately as it turned out) ``unphysical solutions'' and excluded from the quantization. This exclusion is consistent in a free theory since dif\/ferent frequencies decouple there. However, these solutions are perfectly legitimate in the description of local physics. For example, a source outside of the modeled spacetime region will generically generate evanescent waves inside the region (where the theory is source-free by assumption) in addition to propagating ones. In electromagnetism, evanescent waves occur in the near f\/ield of a source.

The (negative of the) symplectic structure (\ref{eq:sympl}) on the space of solutions $L_{x_1}$ associated to the hyperplane at $x_1$ (as in \cite{Oe:KGtl, Oe:timelike} with orientation as a boundary of a region with larger values of $x_1$) is\footnote{See footnote~\ref{fn:nsym}.}
\begin{gather*}
 \omega_{x_1}(\eta_1,\eta_2)   =\frac{1}{2}\int\xd t\,\xd^2 \tilde{x}\,
  \left(\eta_1(t,x_1,\tilde{x}) \partial_1 \eta_2(t,x_1,\tilde{x}) - \eta_2(t,x_1,\tilde{x})\partial_1\eta_1(t,x_1,\tilde{x})\right) \nonumber\\
\hphantom{\omega_{x_1}(\eta_1,\eta_2)}{}
 =\frac{\im}{2}\int\frac{\xd E\,\xd^2 \tilde{k}}{(2\pi)^3 2k_1}
 \left(\eta_2(E,\tilde{k})\overline{\eta_1(E,\tilde{k})}-\overline{\eta_2(E,\tilde{k})}\eta_1(E,\tilde{k})\right) .
\end{gather*}
A compatible complex structure is
\begin{gather*}
 (J(\eta))(E,\tilde{k})=-\im\eta(E,\tilde{k}) .
\end{gather*}
As in the spacelike case this is automatically compatible with evolution in the sense of (\ref{eq:jevol}), now of course in the spatial $x_1$-direction. Thus ``evolution'' in this direction is unitary in the quantum theory, by Proposition~\ref{prop:evol}.

The resulting metric (\ref{eq:kmetric}) is
\begin{gather*}
 g_{x_1}(\eta_1,\eta_2)
= \int\frac{\xd E\,\xd^2 \tilde{k}}{(2\pi)^3 2k_1}
 \left(\eta_1(E,\tilde{k})\overline{\eta_2(E,\tilde{k})}+\eta_2(E,\tilde{k})\overline{\eta_1(E,\tilde{k})}\right) .
\end{gather*}
The complex inner product (\ref{eq:kcip}) is then given by
\begin{gather*}
 \{\eta_1,\eta_2\}_{x_1} = 2\int\frac{\xd E\,\xd^2 \tilde{k}}{(2\pi)^3 2k_1}\,
  \eta_1(E,\tilde{k})\overline{\eta_2(E,\tilde{k})} .
\end{gather*}
The space $L_{x_1}$ is thus precisely the space of (equivalence classes of) complex valued square-integrable functions $\R\times\R^2\to\C$ with the above inner product.

We can split this space into an orthogonal direct sum $L_{x_1}=L_{x_1}^{\text{p}}\oplus L_{x_1}^{\text{e}}$, where the f\/irst summand comprises those functions $\eta\in L_{x_1}$ such that $\eta(E,\tilde{k})=0$ if $E^2-\tilde{k}^2< m^2$ and the second summand comprises those functions $\eta\in L_{x_1}$ such that $\eta(E,\tilde{k})=0$ if $E^2-\tilde{k}^2> m^2$. The elements of $L_{x_1}^{\text{p}}$ are then precisely the \emph{propagating waves}, i.e., those solutions that are globally well behaved. The elements of $L_{x_1}^{\text{e}}$ on the other hand are precisely the \emph{evanescent waves} that behave exponentially in the $x_1$-direction. The orthogonality of the decomposition implies that we obtain a decomposition of the measure on $\hat{L}_{x_1}$ as a product of the measures on the subspaces $\hat{L}^{\text{p}}_{x_1}$ and $\hat{L}^{\text{e}}_{x_1}$. This in turn implies that the respective Hilbert space of square-integrable holomorphic functions decomposes as $\cH_{x_1}=\cH_{x_1}^{\text{p}}\ctens\cH_{x_1}^{\text{e}}$. On the other hand, $\cH_{x_1}^{\text{p}}$ and $\cH_{x_1}^{\text{e}}$ can also be realized as subspaces of $\cH_{x_1}$, namely by restricting to those holomorphic functions that are invariant under translations in $L^{\text{e}}$ or in $L^{\text{p}}$ respectively. It is clear, moreover, that the unitary operator~$U$ associated to evolution in $x_1$-direction by Proposition~\ref{prop:evol} preserves these subspaces. We also note that all coherent states associated to elements of $L_{x_1}^{\text{p}}$ are contained in $\cH_{x_1}^{\text{p}}$ while those associated to elements of $L_{x_1}^{\text{e}}$ are contained in $\cH_{x_1}^{\text{e}}$.

The quantization obtained in \cite{Oe:KGtl, Oe:timelike} can be seen to be equivalent to the restriction of the quantization obtained here to the subspaces $\cH_{x_1}^{\text{p}}$ corresponding to propagating waves only.

\subsection{The hypercylinder}

We consider the following geometrical setting: Admissible hypersurfaces are hypercylinders of the form $\R\times S^2_R$ and their f\/inite disjoint unions, where $S^2_R$ denotes the two-sphere in space of radius $R$, centered at the origin. Admissible regions are all solid hypercylinders of the form $\R\times \bar{B}^3_R$, where $\bar{B}^3_R$ is the closed solid ball of radius $R$ in space centered at the origin. Also admissible are solid hypercylinders with a smaller solid hypercylinder removed. Finite disjoint unions of such regions are also admissible.

We use a cartesian time coordinate $t$ and spherical coordinates $(r,\phi,\theta)$ in space. We also denote the angular coordinates $(\phi,\theta)$ collectively by $\Omega$. $\xd\Omega$ denotes the standard measure on the $2$-sphere of unit radius.
We may parametrize a solution of the Klein--Gordon equation in the neighborhood of a hypercylinder of radius $R$ via functions $\xi:\R\times I\to\C$, where $I=\{(l,m): l\in\N_0, m\in\{-l,-l+1,\dots,l\}\}$, as follows
\begin{gather*}
 \xi(t,r,\Omega)=\int_{-\infty}^{\infty}\xd E\, \frac{p}{4\pi}
  \sum_{l,m}\left(\xi_{l,m}(E) d_l(E,r) e^{-\im E t} Y_l^m(\Omega)
+\overline{\xi_{l,m}(E)}\; \overline{d_l(E,r)} e^{\im E t} Y_l^{-m}(\Omega)\right) .
\end{gather*}
Here $Y_l^m$ denote the spherical harmonics and $p\defeq\sqrt{|E^2-m^2|}$. We also set
\begin{gather*}
 d_l(E,r)\defeq\begin{cases} j_l(p r)+\im n_l(p r) &
  \text{if}\  E^2>m^2,\\
  \im^{-l} j_l(\im p r)-\im^l n_l(\im p r) &
  \text{if}\  E^2<m^2, \end{cases}
\end{gather*}
where $j_l$ and $n_l$ are the spherical Bessel functions of the f\/irst and second kind respectively. Note that the real parts of the functions $d_l$ are chosen to yield solutions that continue to all of the interior of the solid hypercylinder. The modes with $E^2>m^2$ are propagating waves, while the modes with $E^2<m^2$ are evanescent waves.

The (negative of the) symplectic structure (\ref{eq:sympl}) on the space of solutions $L_R$ associated to the hypercylinder of radius $R$, oriented as the boundary of the corresponding solid hypercylinder, is\footnote{See footnote~\ref{fn:nsym}.}
\begin{gather*}
\omega_R(\eta,\xi)   =\frac{R^2}{2}\int \xd t\,\xd\Omega
 \left(\xi(t,R,\Omega) \partial_r \eta(t,R,\Omega)- \eta(t,R,\Omega) \partial_r \xi(t,R,\Omega)\right) \nonumber\\
\hphantom{\omega_R(\eta,\xi)}{}
 = \int_{-\infty}^\infty\xd E\frac{\im p}{8\pi}\sum_{l,m}
 \left(\eta_{l,m}(E)\overline{\xi_{l,m}(E)}-\overline{\eta_{l,m}(E)}\xi_{l,m}(E)\right) .
\end{gather*}
We consider the following complex structure on the space of these solutions
\begin{gather*}
 (J(\xi))_{l,m}(E)=\im\xi_{l,m}(E) .
\end{gather*}
Due to our choice of parametrization in terms of solutions that extend over all of Minkowski space except for the time axis this is automatically compatible with \emph{radial evolution} in the sense of (\ref{eq:jevol}). Thus, we obtain unitary radial evolution in the quantum theory according to Proposition~\ref{prop:evol}.

The metric (\ref{eq:kmetric}) is
\begin{gather*}
 g_R(\eta,\xi) = \int_{-\infty}^\infty\xd E\frac{p}{4\pi}\sum_{l,m}
 \left(\eta_{l,m}(E)\overline{\xi_{l,m}(E)}+\overline{\eta_{l,m}(E)}\xi_{l,m}(E)\right) .
\end{gather*}
The inner product (\ref{eq:kcip}) between solutions is
\begin{gather*}
 \{\eta,\xi\}_R =\int_{-\infty}^\infty\xd E\, \frac{p}{2\pi}\sum_{l,m}
  \overline{\eta_{l,m}(E)} \xi_{l,m}(E) .
\end{gather*}
Thus, the space $L_R$ is the Hilbert space of (equivalence classes of) square-integrable functions $\R\times I\to\C$ with this inner product.

Analogous to the previous Section~\ref{sec:kgtl} we can decompose $L_R$ into an orthogonal direct sum $L_R=L_R^{\text{p}}\oplus L_R^{\text{e}}$, corresponding to radially propagating waves (with $E^2>m^2$) and evanescent waves (with $E^2<m^2$). Correspondingly we obtain subspaces $\cH_R^{\text{p}}$ and $\cH_R^{\text{e}}$ of $\cH_R$ such that $\cH_R=\cH_R^{\text{p}}\ctens\cH_R^{\text{e}}$. Also, radial evolution preserves these subspaces and coherent states behave as expected with respect to the decomposition.

The quantization obtained in \cite{CoOe:spsmatrix,CoOe:smatrixgbf, Oe:KGtl} is equivalent to the quantization obtained here, when restricted to the subspaces $\cH_R^{\text{p}}$. In particular, the normalized coherent state $\tilde{K}_{R,\xi}$ for $\xi\in L_R^{\text{p}}$ takes in the Schr\"odinger representation the form
\begin{gather*}
 \tilde{K}_{R,\xi}(\varphi)= C_{R,\xi}\exp\left(\int\frac{\xd t\,\xd\Omega\,\xd E}{2\pi}
  \sum_{l,m}\frac{\overline{\xi_{l,m}(E)}}{d_l(E, R)}
  e^{\im E t}Y_{l}^{-m}(\Omega)\varphi(t,\Omega)\right)\psi_{R,0}(\varphi) ,
\end{gather*}
which is equation (10) of \cite{CoOe:spsmatrix} and equation (104) of \cite{CoOe:smatrixgbf}, adapted to our present conventions. Observe in particular, that there is a relative complex conjugation of~$\xi_{l,m}(E)$. That is, the state~$\tilde{K}_{R,\xi}$ is the state~$\psi_{R,\xi'}$ in the conventions of \cite{CoOe:spsmatrix,CoOe:smatrixgbf}, where $\xi'_{l,m}(E)=\overline{\xi_{l,m}(E)}$.

Note that in \cite{CoOe:spsmatrix,CoOe:smatrixgbf} interactions were included leading to a mixing of modes. It was thus necessary to extend (the conf\/iguration version of) the space~$L_R^{\text{p}}$ to~$L_R$. In particular, the vacuum was extended to a function on the conf\/iguration space corresponding to all of~$L_R$. However, since the occurring evanescent waves decay exponentially with the radius it was not necessary to include quantizations of them in the asymptotic state space at large radius considered in the $S$-matrix picture developed in~\cite{CoOe:spsmatrix,CoOe:smatrixgbf}.

As a concrete example of how the quantization works out and reproduces corresponding results of~\cite{Oe:KGtl} and~\cite{CoOe:spsmatrix,CoOe:smatrixgbf} we consider a solid hypercylinder of radius $R$. The space $L_R$ of solutions on the boundary decomposes as $L_R=L_{\tilde{R}} \oplus J_R L_{\tilde{R}}$ due to Lemma~\ref{lem:decis}. For an element $\xi_{l,m}(E)=\xi_{l,m}^{\text{R}}(E)+J_R \xi_{l,m}^{\text{I}}(E)$ this takes the form
\begin{gather*}
 \xi^{\text{R}}_{l,m}(E)\defeq\frac{1}{2}\left(\xi_{l,m}(E)+\overline{\xi_{l,-m}(-E)}\right),\\
 \xi^{\text{I}}_{l,m}(E)\defeq-\frac{\im}{2}\left(\xi_{l,m}(E)-\overline{\xi_{l,-m}(-E)}\right) .
\end{gather*}
The decomposition satisf\/ies
\begin{gather*}
 \overline{\xi^{\text{R}}_{l,m}(E)}=\xi^{\text{R}}_{l,-m}(-E),\qquad\overline{\xi^{\text{I}}_{l,m}(E)}=\xi^{\text{I}}_{l,-m}(-E) .
\end{gather*}
The amplitude of a normalized coherent state $\tilde{K}_{R,\xi}$ is then in accordance with equation (\ref{eq:ncohampl})
\begin{gather*}
  \rho\left(\tilde{K}_{R,\xi}\right) =\exp\left(-\frac{1}{2}\{\xi^{\text{I}},\xi^{\text{I}}\}_R-\frac{\im}{2}\{\xi^{\text{R}},\xi^{\text{I}}\}_R\right) ,
\end{gather*}
where both inner products appearing in the exponent are real. This is easily verif\/ied to coincide with equation~(12) of~\cite{CoOe:spsmatrix} and equation~(106) of \cite{CoOe:smatrixgbf}, up to the abovementioned dif\/ference of conventions. We also note that in this context the complex classical solution $\hat{\xi}$ appearing in equation~(14) of \cite{CoOe:spsmatrix} and equation~(124) of~\cite{CoOe:smatrixgbf} is given by $\hat{\xi}=\xi^{\text{R}}-\im\xi^{\text{I}}$.

\section{Discussion and outlook}
\label{sec:outlook}

We start by commenting on some of the distinctive aspects of the results obtained. One permeating feature of the quantization scheme presented here is its manifest locality. This extends in particular to the spaces of classical solutions used as ingredients of the quantization. It is essential that we do not constrain ourselves to ``global'' solutions, for several reasons. The more obvious one is that in the absence of a global background (in the terminology of \cite{Oe:GBQFT}), we do not even know what global solutions should be. Even more important, however, is the interplay between solutions in the interior of a region and on its boundary encoded in axiom~(C5). This means that there is essentially a $2:1$ correspondence between classical solutions on the boundary and in the interior. As explained in Section~\ref{sec:classft}, far from being surprising, this is to be expected. This ingredient plays an absolutely crucial role in the quantization scheme and is key to making axioms~(T3x) and~(T5b) work. Moreover, it is less related to a particular quantization scheme than to the GBF as such, as indicated by similar considerations in the quantization on the hypercylinder in~\cite{Oe:KGtl}, where the quite dif\/ferent Schr\"odinger--Feynman quantization scheme was used.

Another feature of the present quantization scheme that merits some remarks is the fact that amplitudes are def\/ined through an integral (equation (\ref{eq:aint})). This integral is quite dif\/ferent from the Feynman path integral though, in two respects. Firstly, it is an integral over classical solutions only rather than over general f\/ield conf\/igurations. However, in the case of a free theory the Feynman path integral can also be converted to an integral over classical solutions, a fact that has been extensively used for example in \cite{CoOe:unitary,CoOe:spsmatrix,CoOe:smatrixgbf,Oe:KGtl, Oe:timelike}. The second dif\/ference is more decisive: The integral (\ref{eq:aint}) is a proper Gaussian integral, where the exponential has a~real negative def\/inite argument, rather than an imaginary argument as for the Feynman path integral. Note that this fact has nothing to do here with any spacetime metric or its signature.

We turn to a discussion of some directions for future development. An important limitation of the quantization scheme presented here is its restriction to linear f\/ield theories. On the other hand, there exists a considerable body of knowledge on the geometric quantization of non-linear theories \cite{Woo:geomquant}, although much of it is restricted to mechanical systems that are not f\/ield theories. Nevertheless, it is probably not too dif\/f\/icult to transfer some of this to the quantization on hypersurfaces in the present context. What is less clear, however, is how to generalize appropriately the amplitude map (\ref{eq:aint}). It is probable that an integral over classical solutions will not do in this case. Perhaps a ``mixed'' quantization, where some aspects of the Feynman path integral are taken over might be envisaged. To this end it would certainly be desirable to develop a detailed understanding of the relation to other quantization schemes, such as the Schr\"odinger--Feynman one.

Another ``defect'' of the present quantization scheme is its failure to include corners, i.e., hypersurfaces with boundaries. The fact that the axioms of Section~\ref{sec:axioms} dif\/fer little from those presented in~\cite{Oe:2dqym} makes this perhaps surprising. Indeed, no drastic changes might seem to be necessary for the implementation of corners in the present scheme. However, there are some subtle dif\/f\/iculties, which suggest that the form that corners were implemented in the axioms in~\cite{Oe:2dqym} might need to be generalized. In particular, it might be that the state spaces associated to hypersurfaces with boundaries should not in general be required to be complex Hilbert spaces. Notice that this would not be in conf\/lict with the probability interpretation of the GBF~\cite{Oe:GBQFT,Oe:probgbf}. The latter only applies to hypersurfaces that are boundaries and are hence closed. This makes us free to contemplate associating weaker structures to hypersurfaces with boundaries.

One type of object that plays a central role in many quantization schemes, but that we have not mentioned at all so far are observables. This is partly justif\/ied by the fact that one role that observables traditionally play, namely to provide a description of time-evolution, is now served by the amplitude maps. However, observables also play other important roles such as describing particular measurement operations. Although little has been written so far about observables in the GBF, the concepts of observable and expectation value f\/it naturally into the framework \cite{Oe:talk2009pi}. In the present quantization scheme, there is even a natural way to quantize classical observables. Suppose we have a classical observable given by a function $f$ on the space of solutions $L_M$ in a spacetime region $M$. Assuming that $r_M:L_M\to L_{\partial M}$ is injective, we can view $f$ as a function on $L_{\tilde M}=r_M(L_M)$. A quantum observable corresponding to $f$ can then be given by the map $\rho_M^f:\cH_{\partial M}^\ds\to\C$ obtained by inserting $f$ into the integral (\ref{eq:aint}). This amounts to a kind of Berezin--Toeplitz quantization of the observable. While this quantization leads to the expected commutation relations between linear observables in some simple examples it might not be suitable for more complicated observables. In any case these are just some initial remarks and it is clearly necessary to develop the concept of observable in much more depth.

Another subject we have not mentioned here at all is that of symmetries. Due to the functorial nature of the assignment of algebraic to geometric data it is quite natural to consider actions of local spacetime symmetry groups on state spaces and amplitudes \cite{Oe:GBQFT}. Supposing we are given such actions on the local classical spaces of solutions, the present quantization scheme seems particularly amenable to transfer these to actions on the corresponding quantum objects.

Finally, let us mention a point that is perhaps more of interest from the point of view of topological quantum f\/ield theory (TQFT) rather than for the realization of physical theories. As mentioned previously, the integration of the gluing anomaly factor into the axiom (T5b) allows for example to admit regions with non-trivial topology, but without boundaries in the case of certain simple TQFTs with f\/inite-dimensional state spaces. In the inf\/inite-dimensional case the gluing anomaly factor would generically become inf\/inite in such cases. One could now envision some kind of renormalization of this factor. That is, suppose we can ``regularize'' the theory such that the anomaly factor becomes f\/inite depending on some kind of ``cut-of\/f''. If the anomaly factor diverges under removal of the ``cut-of\/f'' in a way that is controlled and compatible with the composition of gluings, this could serve as the def\/inition of a ``renormalized'' theory. In this theory more regions would be admissible than in the original one and the anomaly would contain more data, encoding the precise way in which the anomaly factor diverges.

\subsection*{Acknowledgments}

I would like to thank Daniele Colosi for stimulating
discussions. This work was supported in part by CONACyT grant 49093.

\pdfbookmark[1]{References}{ref}
\LastPageEnding


\begin{thebibliography}{99}
\footnotesize\itemsep=0pt

\bibitem{Ati:tqft}
Atiyah M., Topological quantum f\/ield theories, \textit{Inst. Hautes \'Etudes
  Sci. Publ. Math.}  (1988), 175--186.

\bibitem{Bar:hilbanalytic}
Bargmann V., On a {H}ilbert space of analytic functions and an associated
  integral transform. Part~I, \href{http://dx.doi.org/10.1002/cpa.3160140303}{\textit{Comm. Pure Appl. Math.}} \textbf{14}
  (1961), 187--214.

\bibitem{Bar:remanalytic}
Bargmann V., Remarks on a {H}ilbert space of analytic functions, \href{http://dx.doi.org/10.1073/pnas.48.2.199}{\textit{Proc.
  Nat. Acad. Sci. USA}} \textbf{48} (1962), 199--204.

\bibitem{Boc:analysisprob}
Bochner S., Harmonic analysis and the theory of probability, University of
  California Press, Berkeley and Los Angeles, 1955.

\bibitem{Col:vacuum}
Colosi D., On the structure of the vacuum state in general boundary quantum
  f\/ield theory, \href{http://arxiv.org/abs/0903.2476}{arXiv:0903.2476}.

\bibitem{Col:desitterletter}
Colosi D., $S$-matrix in de~Sitter spacetime from general boundary quantum
  f\/ield theory, \href{http://arxiv.org/abs/0910.2756}{arXiv:0910.2756}.

\bibitem{CoOe:unitary}
Colosi D., Oeckl R.,
On unitary evolution in quantum f\/ield theory in curved spacetime,
\href{http://dx.doi.org/10.2174/1874415X01104010013}{\textit{Open Nuclear Part. Phys.~J.}} \textbf{4} (2011), 13--20,
  \href{http://arxiv.org/abs/0912.0556}{arXiv:0912.0556}.

\bibitem{CoOe:spsmatrix}
Colosi D., Oeckl R., {$S$}-matrix at spatial inf\/inity, \href{http://dx.doi.org/10.1016/j.physletb.2008.06.011}{\textit{Phys. Lett.~B}}
  \textbf{665} (2008), 310--313, \href{http://arxiv.org/abs/0710.5203}{arXiv:0710.5203}.

\bibitem{CoOe:smatrixgbf}
Colosi D., Oeckl R., Spatially asymptotic $S$-matrix from general boundary
  formulation, \href{http://dx.doi.org/10.1103/PhysRevD.78.025020}{\textit{Phys. Rev.~D}} \textbf{78} (2008), 025020, 22~pages,
  \href{http://arxiv.org/abs/0802.2274}{arXiv:0802.2274}.

\bibitem{CoOe:2deucl}
Colosi D., Oeckl R., States and amplitudes for f\/inite regions in a
  two-dimensional {E}uclidean quantum f\/ield theory, \href{http://dx.doi.org/10.1016/j.geomphys.2009.03.004}{\textit{J.~Geom. Phys.}}
  \textbf{59} (2009), 764--780, \href{http://arxiv.org/abs/0811.4166}{arXiv:0811.4166}.

\bibitem{CDORT:vacuum}
Conrady F., Doplicher L., Oeckl R., Rovelli C., Testa M., Minkowski vacuum in
  background independent quantum gravity, \href{http://dx.doi.org/10.1103/PhysRevD.69.064019}{\textit{Phys. Rev.~D}} \textbf{69}
  (2004), 064019, 7~pages, \href{http://arxiv.org/abs/gr-qc/0307118}{gr-qc/0307118}.

\bibitem{CoRo:genschroed}
Conrady F., Rovelli C., Generalized {S}chr\"odinger equation in {E}uclidean
  f\/ield theory, \href{http://dx.doi.org/10.1142/S0217751X04019445}{\textit{Internat.~J. Modern Phys.~A}} \textbf{19} (2004),
  4037--4068, \href{http://arxiv.org/abs/hep-th/0310246}{hep-th/0310246}.

\bibitem{Dop:tomschwing}
Doplicher L., Generalized {T}omonaga--{S}chwinger equation from the {H}adamard
  formula, \href{http://dx.doi.org/10.1103/PhysRevD.70.064037}{\textit{Phys. Rev.~D}} \textbf{70} (2004), 064037, 7~pages,
  \href{http://arxiv.org/abs/gr-qc/0405006}{gr-qc/0405006}.

\bibitem{HaKa:aqft}
Haag R., Kastler D., An algebraic approach to quantum f\/ield theory,
  \href{http://dx.doi.org/10.1063/1.1704187}{\textit{J.~Math. Phys.}} \textbf{5} (1964), 848--861.

\bibitem{Lan:rfanalysis}
Lang S., Real and functional analysis, \href{http://dx.doi.org/10.1007/978-1-4612-0897-6}{\textit{Graduate Texts in Mathematics}},
  Vol.~142, 3rd ed., Springer-Verlag, New York, 1993.

\bibitem{Oe:boundary}
Oeckl R., A ``general boundary'' formulation for quantum mechanics and quantum
  gravity, \href{http://dx.doi.org/10.1016/j.physletb.2003.08.043}{\textit{Phys. Lett.~B}} \textbf{575} (2003), 318--324,
  \href{http://arxiv.org/abs/hep-th/0306025}{hep-th/0306025}.


\bibitem{Oeckl2012}
Oeckl R., Af\/f\/ine holomorphic quantization, \href{http://dx.doi.org/10.1016/j.geomphys.2012.02.001}{\textit{J.~Geom. Phys.}} \textbf{62}
  (2012), 1373--1396, \href{http://arxiv.org/abs/1104.5527}{arXiv:1104.5527}.

\bibitem{Oe:talk2009pi}
Oeckl R., Against commutators, Talk at Perimeter Institute (Waterloo, Canada,
  January 20, 2009), available at \url{http://pirsa.org/09010002/}.

\bibitem{Oe:GBQFT}
Oeckl R., General boundary quantum f\/ield theory: foundations and probability
  interpretation, \textit{Adv. Theor. Math. Phys.} \textbf{12} (2008),
  319--352, \href{http://arxiv.org/abs/hep-th/0509122}{hep-th/0509122}.

\bibitem{Oe:KGtl}
Oeckl R., General boundary quantum f\/ield theory: timelike hypersurfaces in the
  {K}lein--{G}ordon theory, \href{http://dx.doi.org/10.1103/PhysRevD.73.065017}{\textit{Phys. Rev.~D}} \textbf{73} (2006), 065017,
  13~pages, \href{http://arxiv.org/abs/hep-th/0509123}{hep-th/0509123}.

\bibitem{Oe:probgbf}
Oeckl R., Probabilities in the general boundary formulation, \href{http://dx.doi.org/10.1088/1742-6596/67/1/012049}{\textit{J.~Phys.
  Conf. Ser.}} \textbf{67} (2007), 012049, 6~pages, \href{http://arxiv.org/abs/hep-th/0612076}{hep-th/0612076}.

\bibitem{Oe:catandclock}
Oeckl R., Schr\"odinger's cat and the clock: lessons for quantum gravity,
  \href{http://dx.doi.org/10.1088/0264-9381/20/24/009}{\textit{Classical Quantum Gravity}} \textbf{20} (2003), 5371--5380,
  \href{http://arxiv.org/abs/gr-qc/0306007}{gr-qc/0306007}.

\bibitem{Oe:timelike}
Oeckl R., States on timelike hypersurfaces in quantum f\/ield theory,
  \href{http://dx.doi.org/10.1016/j.physletb.2005.06.078}{\textit{Phys. Lett.~B}} \textbf{622} (2005), 172--177,
  \href{http://arxiv.org/abs/hep-th/0505267}{hep-th/0505267}.

\bibitem{Oe:bqgrav}
Oeckl R., The general boundary approach to quantum gravity, in Proceedings of
  the First International Conference on Physics (Tehran, 2004), Amirkabir
  University, Tehran, 2004, 257--265, \href{http://arxiv.org/abs/gr-qc/0312081}{gr-qc/0312081}.

\bibitem{Oe:2dqym}
Oeckl R., Two-dimensional quantum {Y}ang--{M}ills theory with corners,
  \href{http://dx.doi.org/10.1088/1751-8113/41/13/135401}{\textit{J.~Phys.~A: Math. Theor.}} \textbf{41} (2008), 135401, 20~pages,
  \href{http://arxiv.org/abs/hep-th/0608218}{hep-th/0608218}.

\bibitem{OsSc:axeucl}
Osterwalder K., Schrader R., Axioms for {E}uclidean {G}reen's functions,
  \href{http://dx.doi.org/10.1007/BF01645738}{\textit{Comm. Math. Phys.}} \textbf{31} (1973), 83--112.

\bibitem{OsSc:axeucl2}
Osterwalder K., Schrader R., Axioms for {E}uclidean {G}reen's functions.~{II},
  \href{http://dx.doi.org/10.1007/BF01608978}{\textit{Comm. Math. Phys.}} \textbf{42} (1975), 281--305.

\bibitem{Seg:cftdef}
Segal G., The def\/inition of conformal f\/ield theory, in Topology, Geometry and
  Quantum Field Theory, \textit{London Math. Soc. Lecture Note Ser.}, Vol.~308,
  Cambridge University Press, Cambridge, 2004, 421--577.

\bibitem{StWi:pct}
Streater R.F., Wightman A.S., P{CT}, spin and statistics, and all that,
  W.A.~Benjamin, Inc., New York~-- Amsterdam, 1964.

\bibitem{Sud:qintmes}
Sudakov V.N., Linear sets with quasi-invariant measure, \textit{Dokl. Akad.
  Nauk SSSR} \textbf{127} (1959), 524--525.

\bibitem{Tur:qinv}
Turaev V.G., Quantum invariants of knots and 3-manifolds, \textit{de Gruyter
  Studies in Mathematics}, Vol.~18, Walter de Gruyter \& Co., Berlin, 1994.

\bibitem{Wal:qftcurved}
Wald R.M., Quantum f\/ield theory in curved spacetime and black hole
  thermodynamics, \textit{Chicago Lectures in Physics}, University of Chicago Press,
  Chicago, IL, 1994.

\bibitem{Woo:geomquant}
Woodhouse N., Geometric quantization, \textit{Oxford Mathematical Monographs}, The
  Clarendon Press, Oxford University Press, New York, 1980.

\end{thebibliography}
\end{document}